\documentclass[a4paper,aps,prd,10pt,preprintnumbers,showpacs,twocolumn,superscriptaddress,nofootinbib,amsmath,amssymb]{revtex4-1}
\usepackage{graphicx}
\usepackage{cmap}
\usepackage[utf8]{inputenc}
\usepackage[T1]{fontenc}

\def\a{\widetilde{\alpha}}
\def\tx{\tilde{x}}
\def\imo{i}

\begin{document}
\title{Overtones behavior of higher dimensional black holes in the Einstein-Gauss-Bonnet gravity}
\author{Z. Stuchlík}
\email{zdenek.stuchlik@physics.slu.cz}
\affiliation{Research Centre for Theoretical Physics and Astrophysics, Institute of Physics, Silesian University in Opava, Bezručovo náměstí 13, CZ-74601 Opava, Czech Republic}

\author{A. Zhidenko}
\email{olexandr.zhydenko@ufabc.edu.br}
\affiliation{Research Centre for Theoretical Physics and Astrophysics, Institute of Physics, Silesian University in Opava, Bezručovo náměstí 13, CZ-74601 Opava, Czech Republic}
\affiliation{Centro de Matemática, Computação e Cognição (CMCC), Universidade Federal do ABC (UFABC), \\ Rua Abolição, CEP: 09210-180, Santo André, SP, Brazil}

\begin{abstract}
Gravitational perturbations of higher-dimensional black holes in the Einstein-Gauss-Bonnet theory, proposed by Boulware and Deser, have been extensively studied in numerous works, primarily focusing on the fundamental mode. These studies have shown that for sufficiently small black holes, comparing to the Gauss-Bonnet coupling parameter, a dynamical instability arises.
In this work, for the first time, we conduct a comprehensive analysis of the behavior of overtones. We demonstrate that while the fundamental mode remains largely unchanged due to the limited stability region, the first few overtones deviate from their Tangherlini limits at an increasing rate. This deviation reflects the impact of the coupling parameter on the near-horizon structure of the black hole.
\end{abstract}
\pacs{02.60.Lj,04.50.Gh,04.50.Kd,04.70.Bw}
\maketitle

\section{Introduction}

The study of black hole perturbations and quasinormal modes (QNMs) has been a cornerstone of gravitational physics, providing deep insights into the stability and dynamics of black holes in various modified theories of gravity \cite{Konoplya:2011qq,Kokkotas:1999bd,Berti:2009kk,Bolokhov:2025uxz}. In particular, higher-curvature corrections, such as those introduced by the Einstein-Gauss-Bonnet (EGB) gravity, have attracted significant attention in the context of string-theoretic extensions of General Relativity in higher dimensions \cite{Lovelock:1971yv,Boulware:1985wk,Zwiebach:1985uq}.

Gauss-Bonnet gravity, as the simplest extension of General Relativity that incorporates quadratic curvature terms while maintaining second-order field equations, introduces intriguing modifications to black-hole solutions and their associated QNM spectra. The exact static, spherically symmetric black hole solutions in EGB gravity were first derived by Boulware and Deser \cite{Boulware:1985wk}. Their work was later extended to various nonasymptotically flat and charged black holes in \cite{Wheeler:1985qd,Wheeler:1985nh,Wiltshire:1985us,Cai:2001dz}.

While extensive studies have been conducted on the fundamental (dominant) quasinormal modes of various black holes with Gauss-Bonnet corrections \cite{Iyer:1989rd,Konoplya:2004xx,Abdalla:2005hu,Konoplya:2008ix,Chen:2009an,Chen:2015fuf,Konoplya:2020bxa,Yoshida:2015vua,Gonzalez:2017gwa}, relatively little attention has been given to higher overtones -- oscillations with shorter characteristic lifetimes. Unlike fundamental modes, which primarily depend on the geometry of the radiation zone, higher overtones are particularly sensitive to near-horizon deformations. This makes them a valuable tool for probing the near-horizon geometry. Since Gauss-Bonnet gravity introduces higher-curvature corrections that become significant in strong-gravity regimes, it substantially alters the near-horizon structure of black holes. In particular, the Hawking temperature of the Boulware-Deser black hole is significantly lower compared to its counterpart in Einstein gravity, leading to a much longer lifetime for evaporating black holes \cite{Konoplya:2010vz}. Consequently, the higher overtones of Gauss-Bonnet black holes should be particularly sensitive to these corrections, making it possible to distinguish this theory from other alternative gravitational models.

An essential motivation for studying perturbations and quasinormal modes of $D$-dimensional black holes arises from various brane-world models, where the size of the extra dimensions is much larger than the black hole radius \cite{Berti:2003yr,Zinhailo:2024jzt,Kanti:2005xa,Kanti:2004nr,Dubinsky:2024jqi}. In this scenario, the black hole is effectively described by the Tangherlini metric, which means that the two-dimensional spherical element is generalized to a $(D-2)$-dimensional one.

Another crucial factor influencing the quasinormal spectrum is the dynamical eikonal instability, which arises in various theories incorporating higher-curvature corrections, including Einstein-Gauss-Bonnet \cite{Dotti:2005sq,Gleiser:2005ra,Konoplya:2008ix,Cuyubamba:2016cug,Konoplya:2017ymp,Konoplya:2017zwo} and Einstein-Lovelock theories
\cite{Takahashi:2010gz,Takahashi:2011du,Takahashi:2011qda,Konoplya:2017lhs}.
This instability emerges already at relatively small values of the Gauss-Bonnet coupling parameter $\alpha$ and very large multipole numbers $\ell$, leaving little room for variations in the fundamental mode at small $\ell$. However, this limitation does not extend to overtones, which encode valuable information about the near-horizon geometry \cite{Konoplya:2022pbc,Konoplya:2023hqb}. Indeed, numerous recent studies have demonstrated that black holes, which are nearly indistinguishable from the Schwarzschild solution in the far zone but exhibit significant differences in the near-horizon region, have their first few overtones deviating from their Schwarzschild/Kerr limits at an increasingly rapid rate \cite{Konoplya:2023ahd,Konoplya:2023ppx,Konoplya:2022iyn,Konoplya:2025hgp,Konoplya:2024lch,Zhang:2024nny,Bolokhov:2023bwm,Bolokhov:2023ruj,Lutfuoglu:2025ljm,Stashko:2024wuq,Zinhailo:2024kbq}.

Most numerical methods for calculating quasinormal modes rely on approximating the perturbation equations across the entire spacetime, making them particularly suited for determining the dominant modes of the spectrum. The analytic formula describing the asymptotic behavior of higher overtones was derived in \cite{Moura:2022gqm}. Unfortunately, this expression offers a reliable approximation only for highly damped modes and does not enable accurate estimation of the dominant overtones relevant to the ringdown phase. Recently, the overtones of the Boulware-Deser black hole were examined in the context of spectral instability \cite{Cao:2024sot}. However, the spectral method used in these calculations does not allow for an accurate determination of the overtone values. An accurate method, based on Frobenius expansion and capable of computing any mode, was developed by Leaver \cite{Leaver:1985ax}, but it applies only to differential equations with polynomial coefficients. Since the metric function of the Boulware-Deser black hole cannot be reduced to a rational function of the radial coordinate, Leaver's method cannot be directly applied to this case. However, \cite{Konoplya:2023aph} demonstrated that the metric function can be approximated by a rational function using a general parametrization, with the resulting solutions converging to the exact ones.

In this work, we employ rational approximations for the metric function to conduct a detailed investigation of the higher overtones of gravitational perturbations in Gauss-Bonnet black holes, analyzing their behavior as a function of the coupling parameter. We discuss the robustness of the approximation and the convergence properties for different overtone numbers.

The paper is structured as follows. In Sec.~\ref{sec:basic}, we introduce the Einstein-Gauss-Bonnet black hole solutions and the corresponding perturbation equations, providing the necessary theoretical background. In Sec.~\ref{sec:param}, we describe the parametric representation of the Boulware-Deser metric, which enables an efficient approximation of the black hole spacetime. Sec.~\ref{sec:Frobenius} is dedicated to the numerical methods used for calculating quasinormal modes, with a particular focus on the Frobenius method applied to rational approximations of the metric function. In Sec.~\ref{sec:qnms}, we present and analyze our results for the overtones of higher-dimensional Gauss-Bonnet black holes, highlighting their deviations from the Tangherlini values. Finally, in Sec.~\ref{sec:conclusions}, we summarize our findings and discuss possible extensions of this work.

\section{Gauss-Bonnet black holes and perturbation equations}\label{sec:basic}

The Lagrangian of the D-dimensional Einstein-Gauss-Bonnet theory has the form
\begin{equation}\label{gbg3}
\mathcal{L}=-2\Lambda+R+\frac{\alpha}{2}(R_{\mu\nu\lambda\sigma}R^{\mu\nu\lambda\sigma}-4\,R_{\mu\nu}R^{\mu\nu}+R^2).
\end{equation}
In four-dimensional spacetime ($D=4$), the Gauss-Bonnet term reduces to a total divergence, meaning that it does not contribute to the equations of motion. Consequently, its effects become relevant only in higher-dimensional spacetimes. For dimensions $D>6$, additional curvature corrections beyond the second order arise, as described by Lovelock's theory \cite{Lovelock:1971yv}. An exact solution for a static, spherically symmetric black hole in the $D$-dimensional Einstein-Gauss-Bonnet theory (\ref{gbg3}) was first derived by Boulware and Deser in \cite{Boulware:1985wk}. The corresponding metric takes the form
\begin{equation}\label{gbg4}
 ds^2=-f(r)dt^2+\frac{1}{f(r)}dr^2 + r^2\,d\Omega_n^2,
\end{equation}
where $d\Omega_n^2$ is a line element of the $(n=D-2)$-dimensional sphere, and
\begin{equation}\label{fdef}
f(r)=1-r^2\,\psi(r),
\end{equation}
such that it satisfies the following relation:
\begin{equation}\label{Wdef}
W[\psi]\equiv\frac{n}{2}\psi(1 + \a\psi) = \frac{\mu}{r^{n + 1}}\,.
\end{equation}
Here $\mu$ is a constant, proportional to mass, and we introduced the constant proportional the Gauss-Bonnet coupling,
$$\a\equiv \alpha\frac{(n - 1) (n - 2)}{2}.$$
The black hole solution of (\ref{Wdef}), which goes over into the known Tangherlini solutions \cite{Tangherlini:1963bw},
\begin{equation}\label{psidef}
\psi(r)=\frac{1}{2\a}\left(\sqrt{1+\frac{8\a\mu}{nr^{n+1}}}-1\right).
\end{equation}

We shall measure all the quantities in units of the event horizon $r_H$, for which $f(r_H)=0$. We express $\mu$ in terms of $r_H$ as follows:
\begin{equation}\label{massdef}
\mu=\frac{n\,r_H^{n-1}}{2}\left(1+\frac{\a}{r_H^2}\right).
\end{equation}

After decoupling of the angular variables the perturbation equations can be reduced to the second-order master differential equations \cite{Takahashi:2010ye}
\begin{equation}\label{wavelike}
\left(\frac{\partial^2}{\partial t^2}-\frac{\partial^2}{\partial r_*^2}+V_i(r_*)\right)\Psi(t,r_*)=0,
\end{equation}
where $r_*$ is the tortoise coordinate,
\begin{equation}
dr_*\equiv \frac{dr}{f(r)}=\frac{dr}{1-r^2\psi(r)},
\end{equation}
and $i$ stands for $t$ (\emph{tensor}), $v$ (\emph{vector}), and $s$ (\emph{scalar}) perturbations.
The explicit forms of the effective potentials $V_s(r)$, $V_v(r)$, and $V_t(r)$ \cite{Cuyubamba:2016cug} are given by
\begin{eqnarray}\label{potentials}
V_t(r)&=&\frac{\ell(\ell+n-1)f(r)T''(r)}{(n-2)rT'(r)}+\frac{1}{R(r)}\frac{d^2R(r)}{dr_*^2},\\\nonumber
V_v(r)&=&\frac{(\ell-1)(\ell+n)f(r)T'(r)}{(n-1)rT(r)}+R(r)\frac{d^2}{dr_*^2}\Biggr(\frac{1}{R(r)}\Biggr),\\\nonumber
V_s(r)&=&\frac{2\ell(\ell+n-1)f(r)P'(r)}{nrP(r)}+\frac{P(r)}{r}\frac{d^2}{dr_*^2}\left(\frac{r}{P(r)}\right),
\end{eqnarray}
where $\ell=2,3,4,\ldots$ is the multipole number and functions $T(r)$ and $R(r)$ can be written as follows
\begin{eqnarray}
T(r)&=& r^{n-1}\frac{dW}{d\psi}=\frac{nr^{n-1}}{2}\Biggr(1+2\a\psi(r)\Biggr),\\\nonumber
R(r)&=&r\sqrt{T'(r)},\\\nonumber
P(r)&=&\frac{2(\ell-1)(\ell+n)-nr^3\psi'(r)}{\sqrt{T'(r)}}T(r).
\end{eqnarray}

It has been established that gravitational perturbations of asymptotically flat Gauss-Bonnet black holes exhibit instability \cite{Dotti:2005sq,Gleiser:2005ra}. This instability possesses a rather remarkable characteristic from the perspective of its spectral properties \cite{Konoplya:2008ix}: counterintuitively, it develops at high multipole numbers, whereas the lowest multipoles remain stable. This peculiar behavior was later understood to be a consequence of the nonhyperbolicity of the master perturbation equations within the instability region \cite{Reall:2014pwa}. Since the instability is predominantly governed by long-wavelength perturbations, it was named the eikonal instability \cite{Cuyubamba:2016cug}. For the Gauss-Bonnet black holes it has been studied in asymptotically de Sitter and anti-de Sitter spacetimes \cite{Konoplya:2017ymp}. Notably, in the case of asymptotically flat Gauss-Bonnet black holes, stability is ensured for $D=5$, when
\begin{equation}
- r_H^2\frac{3-\sqrt{6}}{6}\leq\a=\alpha\leq r_H^2\frac{\sqrt{2}-1}{2},
\end{equation}
and, for $D=6$, when
\begin{equation}
- r_H^2\frac{1-\sqrt{25-10\sqrt{6}}}{2}\leq\a=3\alpha\leq r_H^2\frac{\sqrt{25+10\sqrt{6}}-1}{2}.
\end{equation}

\section{Parametrized approximation for the Boulware-Deser metric}\label{sec:param}

Here we shall use the general parametrization of a spherically symmetric and asymptotically flat black-hole spacetime by using the continued-fraction expansion in terms of a compact radial coordinate. The parametric ansatz is designed in such a way that the prefactors determine the asymptotic behavior, while the terms in the continued fraction series are fixed by the behavior near the event horizon. This way, the parametrization is valid not only near the black hole or only in the far region, but in the whole space outside the black hole. This idea was first introduced for the four-dimensional spherically symmetric black holes \cite{Rezzolla:2014mua}, extended to axially symmetric black holes \cite{Konoplya:2016jvv} and a class of spherically and axially symmetric wormholes \cite{Bronnikov:2021liv}. Further it was effectively applied in a number works in order to obtain an analytic approximation \cite{Younsi:2016azx,Kokkotas:2017zwt,Kokkotas:2017ymc,Konoplya:2018arm} and further analysis of various properties of four-dimensional black holes in metric theories of gravity \cite{Zinhailo:2019rwd,Konoplya:2019ppy,Nampalliwar:2018iru,Konoplya:2023owh,Konoplya:2022kld,Dubinsky:2024rvf,Dubinsky:2024nzo,Kocherlakota:2020kyu,Zhang:2024rvk,Cassing:2023bpt,Li:2021mnx,Ma:2024kbu,Shashank:2021giy,Konoplya:2021slg,Yu:2021xen,Toshmatov:2023anz,Konoplya:2019fpy,Nampalliwar:2019iti,Ni:2016uik,Zinhailo:2018ska,Paul:2023eep}.

Here we use this parametrization to approximate an exact black hole solution by rational functions, allowing, thereby, for an application of the Frobenius method for finding quasinormal modes. Notice that convergence in the orders of the parametrization must be achieved, which would provide consequent convergence of quasinormal modes \cite{Konoplya:2022iyn,Zinhailo:2018ska}.

This approach was generalized for the higher-dimensional space-times in \cite{Konoplya:2020kqb}. Here we shall use it in order to approximate the metric function of the Boulware-Deser black hole,
\begin{equation}
f(r)\approx\left(1-\frac{r_H^{n - 1}}{r^{n - 1}}\right) A\left(1-\frac{r_H^{n - 1}}{r^{n - 1}}\right)\,,
\label{metAB}
\end{equation}
where the compact coordinate is defined as follows:
\begin{equation}\label{compact}
\tx\equiv1-\frac{r_H^{n - 1}}{r^{n - 1}}\,,
\end{equation}
and $A(\tx)$ is expressed as
\begin{equation}
A(\tx)\equiv 1-\epsilon\,(1-\tx)+(a_0-\epsilon)(1-\tx)^2+{\tilde A}(\tx)(1-\tx)^3\,,
\label{defAx}
\end{equation}
with the coefficients $\epsilon$ and $a_0$, which are fixed by the asymptotic expansion of $f(r)$,
\begin{equation}\label{asymppars}
\epsilon=\frac{\a}{r_H^2},\qquad a_0=0.
\end{equation}

The function ${\tilde A}(\tx)$ is constructed via the continued-fraction expansions,
\begin{equation}
{\tilde A}(\tx)=\frac{a_1}{\displaystyle 1+\frac{\displaystyle a_2\,\tx}{\displaystyle 1+\frac{\displaystyle a_3\,\tx}{\displaystyle 1+\ldots}}}\,.
\end{equation}
and the coefficients $a_1,a_2,a_3,\ldots$ are obtained through the series expansion of $f(r)$ at the event horizon $r=r_H$ ($\tx=0$),
\begin{equation}\nonumber
\begin{array}{rcl}
 a_1 &=& \dfrac{\a \left(4 \a (n-1)+(n-3) r_H^2\right)}{(n-1)r_H^2 \left(2 \a+r_H^2\right)}\,,\\
 a_2 &=& \dfrac{\a (n+1) r_H^2 \left(\a (5 n-3)+2 (n-1) r_H^2\right)}{(n-1) \left(2 \a+r_H^2\right)^2 \left(4 \a (n-1)+(n-3) r_H^2\right)}\\
     &+& \dfrac{(n+1)r_H^4}{2 (n-1) \left(2 \a+r_H^2\right)^2}-\dfrac{3}{2}\,,\ldots.
\end{array}
\end{equation}

The approximation of the metric function is obtained by truncating the continued fraction expansion at order $k$, i.e., by setting $a_{k+1} = 0$. Consequently, the function $f(r)$ is approximated by a rational function of $r$. This simplification reduces the wavelike equation~(\ref{wavelike}) with the effective potentials~(\ref{potentials}) to a second-order differential equation, whose coefficients are polynomials in $r$.

In order to achieve a more accurate approximation for the effective potentials, we eliminate all derivative terms using the relation
\begin{equation}\label{dp}
\psi'(r) = -\frac{(n+1) \psi (r)}{r} \frac{1+\alpha \psi (r)}{1+2 \alpha \psi (r)},
\end{equation}
which follows directly from differentiating Eq.~(\ref{psidef}). Using the definition of the metric function~(\ref{fdef}), we further obtain
\begin{equation}\label{df}
f'(r) = \frac{1-f(r)}{2 r} \left(n-3+\frac{(n+1) r^2}{r^2+2 \alpha-2 \alpha f(r)}\right).
\end{equation}
This expression allows us to construct a more refined approximation for the Eq.~(\ref{wavelike}) in terms of polynomial coefficients.

The derived approximation enables a more detailed and precise analysis of the black-hole spacetime with arbitrary accuracy. By increasing the order of parametrization, we construct a sequence of approximations that rapidly converge to the precise values. The continued fraction approximation exhibits superior convergence properties, leading to significantly more accurate approximations of the metric compared to the series expansion of the function $f(r)$ in (\ref{fdef}) in terms of $\a$. While the latter also provides an approximation in the form of a rational function at any order of the expansion, we shall demonstrate that its convergence is generally slower and less efficient than that of the continued fraction approach.

\section{Numerical method for calculation of quasinormal modes}\label{sec:Frobenius}

Quasinormal modes are the eigenvalues of the wavelike equation~(\ref{wavelike}), subject to the boundary conditions of purely ingoing waves at the event horizon ($r\to r_H$) and purely outgoing waves at spatial infinity ($r\to\infty$). Once the wavelike equation is expressed in terms of polynomial coefficients, the quasinormal spectrum can be determined with high accuracy using Leaver's method~\cite{Leaver:1985ax}. This approach relies on a convergent procedure based on the Frobenius series expansion.

The wavelike equation~(\ref{wavelike}) has a regular singular point at the event horizon $r = r_H$ and an irregular singular point at $r = \infty$. We introduce the new function, $y(r)$, such that
\begin{equation}\label{reg}
\Psi(r) = e^{\imo\omega r} \left(1 - \frac{r_H}{r}\right)^{-i\omega / f'(r_H)} y(r).
\end{equation}
Here $\Psi(r)$ satisfies the quasinormal boundary conditions iff $y(r)$ is regular at $r=r_H$ and $r=\infty$. We express $y(r)$ in terms of the Frobenius series expansion
\begin{equation}\label{Frobenius}
y(r) = \sum_{i=0}^{\infty} a_i \left(1 - \frac{r_H}{r}\right)^i.
\end{equation}
The coefficients $a_i$ satisfy a recurrence relation,
\begin{equation}\label{recurrence}
c_{0,i}a_i+c_{1,i}a_{i-1}+c_{2,i}a_{i-2}+c_{3,i}a_{i-3}+\ldots=0,
\end{equation}
which can be reduced to a three-term recurrence relation via Gaussian elimination (for details, see \cite{Konoplya:2011qq}). The final step involves solving an infinite continued fraction equation for $\omega$, which ensures the convergence of the series~(\ref{Frobenius}) at $r = \infty$.

It is interesting to note that when the approximation (\ref{metAB}) with a truncated expansion at finite order $k$ is used, as described in Sec.~\ref{sec:param}, the analytic structure of the differential equation changes: its singular points correspond to the zeroes of the numerator and denominator of the function $f(r)$ and thus depend on $k$. In particular, $r=0$ becomes an irregular singular point. Depending on the structure of the singularities, the series expansion (\ref{Frobenius}) may or may not converge between the event horizon and infinity. In the latter case, we employ a sequence of positive real midpoints, as outlined in \cite{Rostworowski:2006bp}. Specifically, we choose the first midpoint within the radius of convergence and determine the initial condition at that point using the series expansion, which allows us to construct the new series in the vicinity of the midpoint. This procedure is repeated until the final point $r=\infty$ lies within the radius of convergence. The regularity of the function $y(r)$ for $r_H<r<\infty$ ensures that the result does not depend on the particular choice of midpoints. The recurrence relation in the final interval leads to the equation with infinite continued fraction, for which we use the Nollert improvement \cite{Nollert:1993zz}, generalized in \cite{Zhidenko:2006rs} to recurrence relations with an arbitrary number of terms. It is noteworthy that the number of terms in the recurrence relation varies significantly depending on the of the effective potential (scalar, vector, or tensor type) and the number of spacetime dimensions $D$ in (\ref{potentials}). However, it increases only marginally when improving the approximation order $k$ in the parametrized approximation for the metric function.

Unfortunately, the complexity of the recurrence relations increases rapidly with the number of spacetime dimensions. For example, to accurately compute the quasinormal modes of tensor-type perturbations in $D=6$ with a precision of five decimal places, corresponding to the 13th-order parametrized approximation, we derived a recurrence relation involving 336 terms and employed two midpoints in order to obtain the final equation with infinite continued fraction. Thus, the accurate computation of quasinormal modes in higher dimensions remains a time-consuming and technically demanding task.

\begin{table*}
\begin{tabular}{@{\extracolsep{\fill}}c|@{\hspace{10pt}}c@{\hspace{10pt}}c@{\hspace{10pt}}c@{\hspace{10pt}}c@{\hspace{10pt}}c@{\hspace{10pt}}c}
\multicolumn{2}{c}{$\alpha=-0.09$} & $\alpha=0$ & $\alpha=0.1$ & $\alpha=0.2$ & $\alpha=0.207$ \\
\hline
tensor & $1.42340-0.40511 \imo$ & $1.51057-0.35754 \imo$ & $1.59000-0.34053 \imo$ & $1.64365-0.32874 \imo$ & $1.64657-0.32788 \imo$ \\
 $N=0$ & $1.43646-0.40517 \imo$ & $1.51057-0.35754 \imo$ & $1.59000-0.34053 \imo$ & $1.64366-0.32872 \imo$ & $1.64658-0.32786 \imo$ \\
\hline
tensor & $1.21486-1.28158 \imo$ & $1.39272-1.10456 \imo$ & $1.49059-1.04106 \imo$ & $1.54956-1.00152 \imo$ & $1.55305-0.99893 \imo$ \\
 $N=1$ & $1.28344-1.27611 \imo$ & $1.39273-1.10454 \imo$ & $1.49076-1.04110 \imo$ & $1.55006-1.00142 \imo$ & $1.55353-0.99883 \imo$ \\
\hline
vector & $1.18723-0.32715 \imo$ & $1.13400-0.32752 \imo$ & $1.08850-0.31966 \imo$ & $1.04882-0.30768 \imo$ & $1.04618-0.30679 \imo$ \\
 $N=0$ & $1.18130-0.33659 \imo$ & $1.13761-0.32440 \imo$ & $1.08909-0.31737 \imo$ & $1.04819-0.30664 \imo$ & $1.04552-0.30577 \imo$ \\
\hline
vector & $1.05070-1.02607 \imo$ & $0.94742-1.02204 \imo$ & $0.91912-1.01531 \imo$ & $0.90287-0.97136 \imo$ & $0.90139-0.96793 \imo$ \\
 $N=1$ & $0.90557-1.01638 \imo$ & $0.94939-1.02623 \imo$ & $0.92269-1.02531 \imo$ & $0.92077-0.95888 \imo$ & $0.91393-0.95794 \imo$ \\
\hline
scalar & $1.10209-0.30862 \imo$ & $0.94774-0.25609 \imo$ & $0.82485-0.24552 \imo$ & $0.75517-0.25640 \imo$ & $0.75200-0.25676 \imo$ \\
 $N=0$ & $1.10245-0.30864 \imo$ & $0.95111-0.25888 \imo$ & $0.82370-0.25260 \imo$ & $0.75912-0.26110 \imo$ & $0.75584-0.26104 \imo$ \\
\hline
scalar & $0.97741-0.92217 \imo$ & $0.85123-0.82116 \imo$ & $0.68130-0.76068 \imo$ & $0.59598-0.83048 \imo$ & $0.59751-0.83089 \imo$ \\
 $N=1$ & $0.87519-0.97526 \imo$ & $0.74759-0.85922 \imo$ & $0.57629-0.76810 \imo$ & $0.56223-0.81746 \imo$ & $0.56022-0.81640 \imo$ \\
\end{tabular}
\caption{Dominant mode ($N=0$) and first overtone ($N=1$) of tensor-, vector-, and scalar-type ($\ell=2$) gravitational perturbations for $D=5$ ($r_H=1$). The upper lines show the accurate values obtained using the Frobenius method by taking a sufficiently large approximation order $k$. The lower lines show the results from the 13th-order WKB formula with Padé approximant ($\widetilde{m}=7$, $\widetilde{n}=6$).}\label{tabl:WKBcomparison}
\end{table*}

For quick verification of the obtained results, or when an initial guess is needed, one can employ the higher-order WKB method \cite{Iyer:1986np,Konoplya:2003ii,Konoplya:2004ip} with Padé approximants \cite{Matyjasek:2017psv}. However, unlike the precise Frobenius method, the WKB approach is only asymptotically convergent and yields accurate results only for specific forms of the effective potential -- namely, those with a single peak that can be well approximated by a Taylor expansion between the turning points. In particular, this requires that the overtone number $N$ remains smaller than the multipole number $\ell$. An advantage of the WKB method is that it does not require the metric to be a rational function of the coordinate, allowing us to apply it directly to the accurate metric. Therefore, the WKB method is broadly and effectively used in numerous publications \cite{Konoplya:2020fwg,Konoplya:2021ube,Bolokhov:2023dxq,Skvortsova:2024atk,Malik:2024cgb,Bolokhov:2024ixe,Skvortsova:2024eqi,Lutfuoglu:2025hwh,Dubinsky:2025fwv}. Table~\ref{tabl:WKBcomparison} shows that our precise results agree well with the 13th-order WKB formula with Padé approximant ($\widetilde{m}=7$ and $\widetilde{n}=6$) for $N<\ell$ (see \cite{Konoplya:2019hlu} for details) across all types of perturbations, although the agreement is slightly weaker for scalar-type modes. For higher overtones, $N \geq \ell$, the WKB method is no longer reliable. As the primary aim of this study is the accurate computation of overtones, we do not include WKB results in the present paper.

\section{Quasinormal modes}\label{sec:qnms}

Here, we focus on quasinormal modes with the lowest multipole number, which are typically the dominant ones in the ringdown phase. High multipoles, $\ell \gg n$, have been studied in earlier works \cite{Konoplya:2017wot,Konoplya:2017ymp,Konoplya:2008ix,Konoplya:2004xx}.
It is important to note that when the Gauss-Bonnet correction is introduced, the quasinormal modes of test fields differ from the gravitational ones in the eikonal limit \cite{Konoplya:2017wot}.

\begin{figure}
\resizebox{\linewidth}{!}{\includegraphics{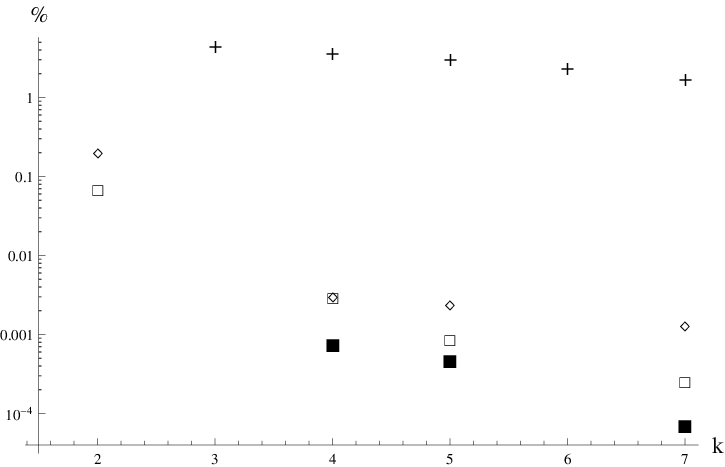}}
\caption{Relative error (in per cent) for the $\ell=2$ modes of the tensor-type perturbations of the $D=5$ Boulware-Deser black hole.
We present the fifth overtone $\alpha=0.207r_H^2$ for various orders the metric function series expansion in $\a$ (pluses) compared to the parametrized approximation of same order $k$ (diamonds).
For comparison we also show the relative errors for other modes: first overtone $\alpha=0.207r_H^2$ (filled squares) and fifth overtone $\alpha=0.1r_H^2$ (empty squares).}\label{fig:convergence}
\end{figure}

\begin{table*}
\begin{tabular}{@{\extracolsep{\fill}}c|@{\hspace{10pt}}c@{\hspace{10pt}}c@{\hspace{10pt}}c@{\hspace{10pt}}c@{\hspace{10pt}}c@{\hspace{10pt}}c}
\multicolumn{2}{c}{$\alpha=-0.09$} & $\alpha=0$ & $\alpha=0.1$ & $\alpha=0.2$ & $\alpha=0.207$ \\
\hline
\hline
 $0$ & $1.42340-~0.40511 \imo$ & $1.51057-0.35754 \imo$ & $1.59000-0.34053 \imo$ & $1.64365-0.32874 \imo$ & $1.64657-0.32788 \imo$ \\
 $1$ & $1.21486-~1.28158 \imo$ & $1.39272-1.10456 \imo$ & $1.49059-1.04106 \imo$ & $1.54956-1.00152 \imo$ & $1.55305-0.99893 \imo$ \\
 $2$ & $0.91417-~2.46242 \imo$ & $1.19383-1.94574 \imo$ & $1.31623-1.79176 \imo$ & $1.35653-1.70861 \imo$ & $1.35890-1.70529 \imo$ \\
 $3$ & $0.87569-~3.73480 \imo$ & $0.99443-2.89898 \imo$ & $1.12022-2.58226 \imo$ & $1.07510-2.28872 \imo$ & $1.06765-2.26599 \imo$ \\
 $4$ & $0.90289-~4.93816 \imo$ & $0.84596-3.91466 \imo$ & $0.98409-3.36821 \imo$ & $1.15498-2.88174 \imo$ & $1.16912-2.85539 \imo$ \\
 $5$ & $0.94962-~6.10906 \imo$ & $0.74362-4.94827 \imo$ & $0.94998-4.19051 \imo$ & $1.18597-3.71366 \imo$ & $1.19770-3.68739 \imo$ \\
 $6$ & $1.00497-~7.26075 \imo$ & $0.67115-5.98316 \imo$ & $0.94672-5.07433 \imo$ & $1.17203-4.55709 \imo$ & $1.18242-4.52635 \imo$ \\
 $7$ & $1.06462-~8.39977 \imo$ & $0.61744-7.01501 \imo$ & $0.93911-5.98520 \imo$ & $1.14670-5.40103 \imo$ & $1.15618-5.36510 \imo$ \\
 $8$ & $1.12654-~9.52974 \imo$ & $0.57596-8.04311 \imo$ & $0.92570-6.90550 \imo$ & $1.11680-6.24385 \imo$ & $1.12558-6.20244 \imo$ \\
 $9$ & $1.18967-10.65275 \imo$ & $0.54287-9.06777 \imo$ & $0.90838-7.82953 \imo$ & $1.08487-7.08487 \imo$ & $1.09308-7.03783 \imo$
\end{tabular}
\caption{Quasinormal modes (fundamental mode and first nine overtones) of the tensor-type ($\ell=2$) gravitational perturbations for $D=5$ ($r_H=1$).}\label{tabl:tensortype}
\end{table*}

\begin{table*}
\begin{tabular}{@{\extracolsep{\fill}}c|@{\hspace{10pt}}c@{\hspace{10pt}}c@{\hspace{10pt}}c@{\hspace{10pt}}c@{\hspace{10pt}}c@{\hspace{10pt}}c}
\multicolumn{2}{c}{$\alpha=-0.09$} & $\alpha=0$ & $\alpha=0.1$ & $\alpha=0.2$ & $\alpha=0.207$ \\
\hline
\hline
 $0$ & $1.18723-0.32715 \imo$ & $1.13400-0.32752 \imo$ & $1.08850-0.31966 \imo$ & $1.04882-0.30768 \imo$ & $1.04618-0.30679 \imo$ \\
 $1$ & $1.05070-1.02607 \imo$ & $0.94742-1.02204 \imo$ & $0.91912-1.01531 \imo$ & $0.90287-0.97136 \imo$ & $0.90139-0.96793 \imo$ \\
 $2$ & $0.84169-1.90071 \imo$ & $0.54293-1.92466 \imo$ & $0.70224-1.88583 \imo$ & $0.70384-1.74527 \imo$ & $0.70243-1.73639 \imo$ \\
 $3$ & $0.70066-2.98715 \imo$ & $0.43570-3.12093 \imo$ & $0.56679-2.80549 \imo$ & $0.52367-2.55333 \imo$ & $0.51901-2.53813 \imo$ \\
 $4$ & $0.66676-4.14336 \imo$ & $0.39958-4.17984 \imo$ & $0.46240-3.70960 \imo$ & $0.34569-3.34840 \imo$ & $0.33364-3.32907 \imo$ \\
 $5$ & $0.68781-5.29445 \imo$ & $0.37154-5.21684 \imo$ & $0.37296-4.59444 \imo$ & $0.24397-4.35030 \imo$ & $0.26810-4.32699 \imo$ \\
 $6$ & $0.73343-6.42852 \imo$ & $0.35018-6.24408 \imo$ & $0.29142-5.47194 \imo$ & $0.35753-5.19198 \imo$ & $0.36590-5.15774 \imo$ \\
 $7$ & $0.79236-7.54831 \imo$ & $0.33355-7.26539 \imo$ & $0.17909-6.37080 \imo$ & $0.39213-6.02610 \imo$ & $0.39183-5.98795 \imo$ \\
 $8$ & $0.85959-8.65894 \imo$ & $0.32024-8.28267 \imo$ & $0.16712-7.56178 \imo$ & $0.37018-6.86941 \imo$ & $0.35758-6.82923 \imo$ \\
 $9$ & $0.93183-9.76428 \imo$ & $0.30934-9.29704 \imo$ & $0.35965-9.40171 \imo$ & $0.27547-8.82903 \imo$ & $0.29743-8.78328 \imo$
\end{tabular}
\caption{Quasinormal modes (fundamental mode and first nine overtones) of the vector-type ($\ell=2$) gravitational perturbations for $D=5$ ($r_H=1$).}\label{tabl:vectortype}
\end{table*}

\begin{table*}
\begin{tabular}{@{\extracolsep{\fill}}c|@{\hspace{10pt}}c@{\hspace{10pt}}c@{\hspace{10pt}}c@{\hspace{10pt}}c@{\hspace{10pt}}c@{\hspace{10pt}}c}
\multicolumn{2}{c}{$\alpha=-0.09$} & $\alpha=0$ & $\alpha=0.1$ & $\alpha=0.2$ & $\alpha=0.207$ \\
\hline
\hline
 $0$ & $ 1.10209-0.30862\imo$ & $ 0.94774-0.25609\imo$ & $ 0.82485-0.24552\imo$ & $ 0.75517-0.25640\imo$ & $ 0.75200-0.25676\imo$ \\
 $1$ & $ 0.97741-0.92217\imo$ & $ 0.85123-0.82116\imo$ & $ 0.68130-0.76068\imo$ & $ 0.59598-0.83048\imo$ & $ 0.59751-0.83089\imo$ \\
 $2$ & $ 0.83569-1.73554\imo$ & $ 0.67273-1.54307\imo$ & $ 0.41393-1.38583\imo$ & $ 0.41938-1.56250\imo$ & $ 0.42387-1.55553\imo$ \\
 $3$ & $ 0.67193-2.79092\imo$ & $ 0.51213-2.43986\imo$ & $ 0.00334-2.13956\imo$ & $ 0.27615-2.36023\imo$ & $ 0.27660-2.34950\imo$ \\
 $4$ & $ 0.48898-3.94582\imo$ & $ 0.57895-3.21851\imo$ & $ 0.08332-3.37833\imo$ & $ 0.04204-3.01002\imo$ & $ 0.02862-2.97764\imo$ \\
 $5$ & $ 0.37354-5.48688\imo$ & $ 0.36192-4.40916\imo$ & $ 0.09756-4.20522\imo$ & $ 0.19445-3.55717\imo$ & $ 0.20156-3.52446\imo$ \\
 $6$ & $ 0.54733-6.65583\imo$ & $ 0.32754-5.41021\imo$ & $ 0.15115-5.00559\imo$ & $ 0.28583-4.38292\imo$ & $ 0.28125-4.35279\imo$ \\
 $7$ & $ 0.62355-7.77251\imo$ & $ 0.30430-6.41305\imo$ & $ 0.23370-5.84750\imo$ & $ 0.25869-5.22828\imo$ & $ 0.22715-5.19889\imo$ \\
 $8$ & $ 0.67908-8.85284\imo$ & $ 0.28762-7.41637\imo$ & $ 0.30102-6.72602\imo$ & $ 0.23943-6.36305\imo$ & $ 0.29034-6.32670\imo$ \\
 $9$ & $ 0.74677-9.90656\imo$ & $ 0.27507-8.41970\imo$ & $ 0.34555-7.62088\imo$ & $ 0.37627-7.21107\imo$ & $ 0.39214-7.16641\imo$
\end{tabular}
\caption{Quasinormal modes (fundamental mode and first nine overtones) of the scalar-type ($\ell=2$) gravitational perturbations for $D=5$ ($r_H=1$).}\label{tabl:scalartype}
\end{table*}

Using the parametrized approximation of the metric function, we can compute not only the dominant quasinormal mode but also the overtones with arbitrary accuracy. Namely, we increase the approximation order $k$ until the desired tolerance is achieved. Due to the superior convergence properties of the continued fraction, the relative error is already within fractions of a percent for the second-order approximation and decreases rapidly as higher-order terms are taken into account. However, it is important to note that truncating the continued fraction at a fixed order is not always possible due to the emergence of singularities in the approximate metric function. In such cases, we bypass the problematic order and instead consider a higher-order parametrized approximation that remains free of these singularities (see Sec. IV of \cite{Konoplya:2016jvv} for a detailed discussion on this issue).

Figure~\ref{fig:convergence} shows the relative error as a function of the approximation order $k$. It is evident that the error associated with the parametrized approximation is several orders of magnitude smaller than that of the series expansion. Although the error increases with both the coupling parameter and the overtone number, it rapidly decreases as $k$ grows. This allows the parametrized approximation to provide accurate approximations even for high overtones and large values of the coupling parameter. In practice, to achieve an accuracy of five decimal places, we typically employ approximations of orders ranging from 8 to 11.

\begin{figure}
\resizebox{\linewidth}{!}{\includegraphics{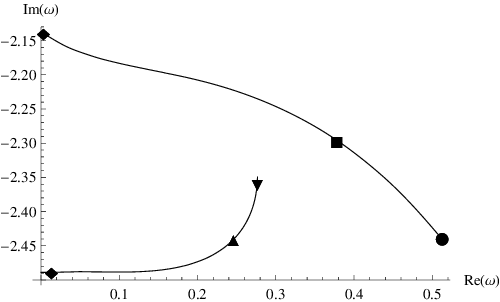}}
\caption{Third overtone of $\ell=2$ scalar-type perturbations for the $D=5$ Boulware-Deser black hole ($r_H=1$). Markers indicate different values of the Gauss-Bonnet coupling: circle for $\alpha = 0$, square for $\alpha = 0.05$, diamonds for $\alpha = 0.1$ and $\alpha = 0.103$, upward triangle for $\alpha = 0.15$, and downward triangle for $\alpha = 0.2$.}\label{fig:d5l2n3scalar}
\end{figure}

Tables~\ref{tabl:tensortype},~\ref{tabl:vectortype},~\ref{tabl:scalartype} present the 10 dominant quasinormal modes of the $D=5$ Boulware-Deser black hole for various coupling parameters within the stability region for gravitational perturbations. While the fundamental mode exhibits an almost linear dependence on the coupling parameter, changing its value only in the second decimal place within the stability region (cf.~\cite{Konoplya:2008ix}), the overtones demonstrate significantly higher sensitivity to variations in $\alpha$. In particular, for scalar-type perturbations (Table~\ref{tabl:scalartype}), we observe that, similar to the behavior found in four-dimensional regular black holes~\cite{Konoplya:2022hll}, the real part of the third and higher overtones undergoes substantial changes within the stability region, decreasing to very small values for $\alpha\simeq 0.1r_H^2$ (see Fig.~\ref{fig:d5l2n3scalar}). This pronounced sensitivity, referred to as the ``outburst of the overtones'' \cite{Konoplya:2023aph}, has been observed not only for various regular black holes (cf.~Fig.~7 of \cite{Konoplya:2022hll}) but also for asymptotically AdS black holes \cite{Konoplya:2023kem}.

\begin{table*}
\begin{tabular}{@{\extracolsep{\fill}}c|@{\hspace{5pt}}c@{\hspace{5pt}}c@{\hspace{5pt}}c@{\hspace{5pt}}c@{\hspace{5pt}}c@{\hspace{5pt}}c}
\multicolumn{2}{c}{$\alpha=0$} & $\alpha=0.1$  & $\alpha=0.2$ & $\alpha=0.3$ & $\alpha=0.4$ & $\alpha=0.5$ \\
\hline
\hline
 $0$ & $ 2.01153-0.50194\imo$ & $2.01116-0.45208 \imo$ & $1.95569-0.42683 \imo$ & $1.88811-0.40857 \imo$ & $1.82176-0.39420 \imo$ & $1.76037-0.38229 \imo$ \\
 $1$ & $ 1.78604-1.55879\imo$ & $1.81994-1.37711 \imo$ & $1.76007-1.29818 \imo$ & $1.69544-1.25274 \imo$ & $1.63838-1.22016 \imo$ & $1.58889-1.19255 \imo$ \\
 $2$ & $ 1.37829-2.81127\imo$ & $1.45992-2.34658 \imo$ & $1.30932-2.12222 \imo$ & $1.27254-2.55802 \imo$ & $1.34158-2.41223 \imo$ & $1.36347-2.31175 \imo$ \\
 $3$ & $ 1.46779-3.63182\imo$ & $1.11684-3.32896 \imo$ & $1.15113-2.82204 \imo$ & $1.28279-3.78762 \imo$ & $1.33598-3.58810 \imo$ & $1.36126-3.42942 \imo$ \\
 $4$ & $ 1.00617-4.33625\imo$ & $0.95492-4.48711 \imo$ & $1.18280-4.08440 \imo$ & $1.30219-5.02053 \imo$ & $1.37068-4.74258 \imo$ & $1.40954-4.52161 \imo$ \\
 $5$ & $ 0.79868-5.94983\imo$ & $0.82540-5.98331 \imo$ & $1.16649-5.38002 \imo$ & $1.35309-6.22606 \imo$ & $1.44031-5.86745 \imo$ & $1.48390-5.59048 \imo$
\end{tabular}
\caption{Quasinormal modes (fundamental mode and first five overtones) of the tensor-type ($\ell=2$) gravitational perturbations for $D=6$ ($r_H=1$).}\label{tabl:d6tensortype}
\end{table*}

In Table~\ref{tabl:d6tensortype} we give several dominant modes for the tensor-type gravitational perturbations for $D=6$. We observe that even the fundamental quasinormal modes display a clear nonlinear dependence of both their real and imaginary parts on the Gauss-Bonnet coupling parameter $\alpha$. Furthermore, higher overtones show an even greater sensitivity to variations in $\alpha$, with the real part of the frequency undergoing substantial changes even for relatively small values of the coupling.

An analogous trend of increasing sensitivity to the Gauss-Bonnet coupling $\alpha$ is also observed in the regime of asymptotically high overtones, as demonstrated using the monodromy method in \cite{Moura:2022gqm}. Remarkably, even for arbitrarily small values of $\alpha$, either positive or negative, the asymptotic quasinormal mode formula acquires a complex multiplicative factor due to the Gauss-Bonnet term. As a result, the real part of the quasinormal frequencies does not approach a constant as it does in the case of the Schwarzschild-Tangherlini black hole, but instead grows without bound. Therefore, we can conclude that for any fixed $\alpha$, there exist overtones whose real parts become arbitrarily larger than the corresponding values for the Tangherlini black hole. Although the data we present lie outside the regime of asymptotically high overtones, we already observe an increase in the real part beginning at some overtone number. This growth sets in at lower overtones for larger values of $\alpha$, which is in agreement with the analytic formula \cite{Moura:2022gqm}.

\section{Conclusions}\label{sec:conclusions}

Higher-dimensional black holes in the Einstein-Gauss-Bonnet theory proposed by Boulware and Deser play an important role in the description of quantum corrected black holes inspired by string theory at low energies. The black-hole metric in this theory suffers from instabilities unless the coupling constant is sufficiently small. While the fundamental mode was thoroughly studied in the literature, the first several overtones, which bring information on the geometry near the event horizon \cite{Konoplya:2022pbc}, have not been studied. At the same time these overtones are necessary to reproduce the early ringdown phase \cite{Giesler:2019uxc,Giesler:2024hcr}. Using the precise Frobenius method, we have studied the overtones' behavior in the stability sector.

We have shown that while the fundamental mode at the lowest multipole numbers changes only slightly due to the narrow parametric range of stability determined by the onset of the eikonal instability, the overtones deviate from their Tangherlini values at an increasing rate. This reflects the fact that the Gauss-Bonnet correction primarily deforms the geometry in the near-horizon region.

The application of the Frobenius method for accurate calculation of the overtones requires the metric function to be a rational function of the radial coordinate. This can be achieved either through a Taylor expansion of the metric in terms of the small coupling parameter or via the parametrized approximation of the metric function by continued fractions. We explored both approaches and demonstrated that the parametrization converges significantly faster, thereby serving as an effective tool for systematically approximating the black hole metric throughout the entire space outside the event horizon.

Our work can be extended to include higher-order corrections, such as those arising from Lovelock theory, which contributes nontrivially to the black hole spacetime in dimensions higher than six.

\bibliography{bibliography}

\begin{thebibliography}{106}%
\makeatletter
\providecommand \@ifxundefined [1]{%
 \@ifx{#1\undefined}
}%
\providecommand \@ifnum [1]{%
 \ifnum #1\expandafter \@firstoftwo
 \else \expandafter \@secondoftwo
 \fi
}%
\providecommand \@ifx [1]{%
 \ifx #1\expandafter \@firstoftwo
 \else \expandafter \@secondoftwo
 \fi
}%
\providecommand \natexlab [1]{#1}%
\providecommand \enquote  [1]{``#1''}%
\providecommand \bibnamefont  [1]{#1}%
\providecommand \bibfnamefont [1]{#1}%
\providecommand \citenamefont [1]{#1}%
\providecommand \href@noop [0]{\@secondoftwo}%
\providecommand \href [0]{\begingroup \@sanitize@url \@href}%
\providecommand \@href[1]{\@@startlink{#1}\@@href}%
\providecommand \@@href[1]{\endgroup#1\@@endlink}%
\providecommand \@sanitize@url [0]{\catcode `\\12\catcode `\$12\catcode
  `\&12\catcode `\#12\catcode `\^12\catcode `\_12\catcode `\%12\relax}%
\providecommand \@@startlink[1]{}%
\providecommand \@@endlink[0]{}%
\providecommand \url  [0]{\begingroup\@sanitize@url \@url }%
\providecommand \@url [1]{\endgroup\@href {#1}{\urlprefix }}%
\providecommand \urlprefix  [0]{URL }%
\providecommand \Eprint [0]{\href }%
\providecommand \doibase [0]{http://dx.doi.org/}%
\providecommand \selectlanguage [0]{\@gobble}%
\providecommand \bibinfo  [0]{\@secondoftwo}%
\providecommand \bibfield  [0]{\@secondoftwo}%
\providecommand \translation [1]{[#1]}%
\providecommand \BibitemOpen [0]{}%
\providecommand \bibitemStop [0]{}%
\providecommand \bibitemNoStop [0]{.\EOS\space}%
\providecommand \EOS [0]{\spacefactor3000\relax}%
\providecommand \BibitemShut  [1]{\csname bibitem#1\endcsname}%
\let\auto@bib@innerbib\@empty
\bibitem [{\citenamefont {Konoplya}\ and\ \citenamefont
  {Zhidenko}(2011)}]{Konoplya:2011qq}%
  \BibitemOpen
  \bibfield  {author} {\bibinfo {author} {\bibfnamefont {R.~A.}\ \bibnamefont
  {Konoplya}}\ and\ \bibinfo {author} {\bibfnamefont {A.}~\bibnamefont
  {Zhidenko}},\ }\href {\doibase 10.1103/RevModPhys.83.793} {\bibfield
  {journal} {\bibinfo  {journal} {Rev. Mod. Phys.}\ }\textbf {\bibinfo {volume}
  {83}},\ \bibinfo {pages} {793} (\bibinfo {year} {2011})},\ \Eprint
  {http://arxiv.org/abs/1102.4014} {arXiv:1102.4014 [gr-qc]} \BibitemShut
  {NoStop}%
\bibitem [{\citenamefont {Kokkotas}\ and\ \citenamefont
  {Schmidt}(1999)}]{Kokkotas:1999bd}%
  \BibitemOpen
  \bibfield  {author} {\bibinfo {author} {\bibfnamefont {K.~D.}\ \bibnamefont
  {Kokkotas}}\ and\ \bibinfo {author} {\bibfnamefont {B.~G.}\ \bibnamefont
  {Schmidt}},\ }\href {\doibase 10.12942/lrr-1999-2} {\bibfield  {journal}
  {\bibinfo  {journal} {Living Rev. Rel.}\ }\textbf {\bibinfo {volume} {2}},\
  \bibinfo {pages} {2} (\bibinfo {year} {1999})},\ \Eprint
  {http://arxiv.org/abs/gr-qc/9909058} {arXiv:gr-qc/9909058} \BibitemShut
  {NoStop}%
\bibitem [{\citenamefont {Berti}\ \emph {et~al.}(2009)\citenamefont {Berti},
  \citenamefont {Cardoso},\ and\ \citenamefont {Starinets}}]{Berti:2009kk}%
  \BibitemOpen
  \bibfield  {author} {\bibinfo {author} {\bibfnamefont {E.}~\bibnamefont
  {Berti}}, \bibinfo {author} {\bibfnamefont {V.}~\bibnamefont {Cardoso}}, \
  and\ \bibinfo {author} {\bibfnamefont {A.~O.}\ \bibnamefont {Starinets}},\
  }\href {\doibase 10.1088/0264-9381/26/16/163001} {\bibfield  {journal}
  {\bibinfo  {journal} {Class. Quant. Grav.}\ }\textbf {\bibinfo {volume}
  {26}},\ \bibinfo {pages} {163001} (\bibinfo {year} {2009})},\ \Eprint
  {http://arxiv.org/abs/0905.2975} {arXiv:0905.2975 [gr-qc]} \BibitemShut
  {NoStop}%
\bibitem [{\citenamefont {Bolokhov}\ and\ \citenamefont
  {Skvortsova}(2025)}]{Bolokhov:2025uxz}%
  \BibitemOpen
  \bibfield  {author} {\bibinfo {author} {\bibfnamefont {S.~V.}\ \bibnamefont
  {Bolokhov}}\ and\ \bibinfo {author} {\bibfnamefont {M.}~\bibnamefont
  {Skvortsova}},\ }\href@noop {} {\  (\bibinfo {year} {2025})},\ \Eprint
  {http://arxiv.org/abs/2504.05014} {arXiv:2504.05014 [gr-qc]} \BibitemShut
  {NoStop}%
\bibitem [{\citenamefont {Lovelock}(1971)}]{Lovelock:1971yv}%
  \BibitemOpen
  \bibfield  {author} {\bibinfo {author} {\bibfnamefont {D.}~\bibnamefont
  {Lovelock}},\ }\href {\doibase 10.1063/1.1665613} {\bibfield  {journal}
  {\bibinfo  {journal} {J. Math. Phys.}\ }\textbf {\bibinfo {volume} {12}},\
  \bibinfo {pages} {498} (\bibinfo {year} {1971})}\BibitemShut {NoStop}%
\bibitem [{\citenamefont {Boulware}\ and\ \citenamefont
  {Deser}(1985)}]{Boulware:1985wk}%
  \BibitemOpen
  \bibfield  {author} {\bibinfo {author} {\bibfnamefont {D.~G.}\ \bibnamefont
  {Boulware}}\ and\ \bibinfo {author} {\bibfnamefont {S.}~\bibnamefont
  {Deser}},\ }\href {\doibase 10.1103/PhysRevLett.55.2656} {\bibfield
  {journal} {\bibinfo  {journal} {Phys. Rev. Lett.}\ }\textbf {\bibinfo
  {volume} {55}},\ \bibinfo {pages} {2656} (\bibinfo {year}
  {1985})}\BibitemShut {NoStop}%
\bibitem [{\citenamefont {Zwiebach}(1985)}]{Zwiebach:1985uq}%
  \BibitemOpen
  \bibfield  {author} {\bibinfo {author} {\bibfnamefont {B.}~\bibnamefont
  {Zwiebach}},\ }\href {\doibase 10.1016/0370-2693(85)91616-8} {\bibfield
  {journal} {\bibinfo  {journal} {Phys. Lett. B}\ }\textbf {\bibinfo {volume}
  {156}},\ \bibinfo {pages} {315} (\bibinfo {year} {1985})}\BibitemShut
  {NoStop}%
\bibitem [{\citenamefont {Wheeler}(1986{\natexlab{a}})}]{Wheeler:1985qd}%
  \BibitemOpen
  \bibfield  {author} {\bibinfo {author} {\bibfnamefont {J.~T.}\ \bibnamefont
  {Wheeler}},\ }\href {\doibase 10.1016/0550-3213(86)90388-3} {\bibfield
  {journal} {\bibinfo  {journal} {Nucl. Phys. B}\ }\textbf {\bibinfo {volume}
  {273}},\ \bibinfo {pages} {732} (\bibinfo {year}
  {1986}{\natexlab{a}})}\BibitemShut {NoStop}%
\bibitem [{\citenamefont {Wheeler}(1986{\natexlab{b}})}]{Wheeler:1985nh}%
  \BibitemOpen
  \bibfield  {author} {\bibinfo {author} {\bibfnamefont {J.~T.}\ \bibnamefont
  {Wheeler}},\ }\href {\doibase 10.1016/0550-3213(86)90268-3} {\bibfield
  {journal} {\bibinfo  {journal} {Nucl. Phys. B}\ }\textbf {\bibinfo {volume}
  {268}},\ \bibinfo {pages} {737} (\bibinfo {year}
  {1986}{\natexlab{b}})}\BibitemShut {NoStop}%
\bibitem [{\citenamefont {Wiltshire}(1986)}]{Wiltshire:1985us}%
  \BibitemOpen
  \bibfield  {author} {\bibinfo {author} {\bibfnamefont {D.~L.}\ \bibnamefont
  {Wiltshire}},\ }\href {\doibase 10.1016/0370-2693(86)90681-7} {\bibfield
  {journal} {\bibinfo  {journal} {Phys. Lett. B}\ }\textbf {\bibinfo {volume}
  {169}},\ \bibinfo {pages} {36} (\bibinfo {year} {1986})}\BibitemShut
  {NoStop}%
\bibitem [{\citenamefont {Cai}(2002)}]{Cai:2001dz}%
  \BibitemOpen
  \bibfield  {author} {\bibinfo {author} {\bibfnamefont {R.-G.}\ \bibnamefont
  {Cai}},\ }\href {\doibase 10.1103/PhysRevD.65.084014} {\bibfield  {journal}
  {\bibinfo  {journal} {Phys. Rev. D}\ }\textbf {\bibinfo {volume} {65}},\
  \bibinfo {pages} {084014} (\bibinfo {year} {2002})},\ \Eprint
  {http://arxiv.org/abs/hep-th/0109133} {arXiv:hep-th/0109133} \BibitemShut
  {NoStop}%
\bibitem [{\citenamefont {Iyer}\ \emph {et~al.}(1989)\citenamefont {Iyer},
  \citenamefont {Iyer},\ and\ \citenamefont {Vishveshwara}}]{Iyer:1989rd}%
  \BibitemOpen
  \bibfield  {author} {\bibinfo {author} {\bibfnamefont {B.~R.}\ \bibnamefont
  {Iyer}}, \bibinfo {author} {\bibfnamefont {S.}~\bibnamefont {Iyer}}, \ and\
  \bibinfo {author} {\bibfnamefont {C.~V.}\ \bibnamefont {Vishveshwara}},\
  }\href {\doibase 10.1088/0264-9381/6/11/016} {\bibfield  {journal} {\bibinfo
  {journal} {Class. Quant. Grav.}\ }\textbf {\bibinfo {volume} {6}},\ \bibinfo
  {pages} {1627} (\bibinfo {year} {1989})}\BibitemShut {NoStop}%
\bibitem [{\citenamefont {Konoplya}(2005)}]{Konoplya:2004xx}%
  \BibitemOpen
  \bibfield  {author} {\bibinfo {author} {\bibfnamefont {R.}~\bibnamefont
  {Konoplya}},\ }\href {\doibase 10.1103/PhysRevD.71.024038} {\bibfield
  {journal} {\bibinfo  {journal} {Phys. Rev. D}\ }\textbf {\bibinfo {volume}
  {71}},\ \bibinfo {pages} {024038} (\bibinfo {year} {2005})},\ \Eprint
  {http://arxiv.org/abs/hep-th/0410057} {arXiv:hep-th/0410057} \BibitemShut
  {NoStop}%
\bibitem [{\citenamefont {Abdalla}\ \emph {et~al.}(2005)\citenamefont
  {Abdalla}, \citenamefont {Konoplya},\ and\ \citenamefont
  {Molina}}]{Abdalla:2005hu}%
  \BibitemOpen
  \bibfield  {author} {\bibinfo {author} {\bibfnamefont {E.}~\bibnamefont
  {Abdalla}}, \bibinfo {author} {\bibfnamefont {R.~A.}\ \bibnamefont
  {Konoplya}}, \ and\ \bibinfo {author} {\bibfnamefont {C.}~\bibnamefont
  {Molina}},\ }\href {\doibase 10.1103/PhysRevD.72.084006} {\bibfield
  {journal} {\bibinfo  {journal} {Phys. Rev. D}\ }\textbf {\bibinfo {volume}
  {72}},\ \bibinfo {pages} {084006} (\bibinfo {year} {2005})},\ \Eprint
  {http://arxiv.org/abs/hep-th/0507100} {arXiv:hep-th/0507100} \BibitemShut
  {NoStop}%
\bibitem [{\citenamefont {Konoplya}\ and\ \citenamefont
  {Zhidenko}(2008)}]{Konoplya:2008ix}%
  \BibitemOpen
  \bibfield  {author} {\bibinfo {author} {\bibfnamefont {R.~A.}\ \bibnamefont
  {Konoplya}}\ and\ \bibinfo {author} {\bibfnamefont {A.}~\bibnamefont
  {Zhidenko}},\ }\href {\doibase 10.1103/PhysRevD.77.104004} {\bibfield
  {journal} {\bibinfo  {journal} {Phys. Rev. D}\ }\textbf {\bibinfo {volume}
  {77}},\ \bibinfo {pages} {104004} (\bibinfo {year} {2008})},\ \Eprint
  {http://arxiv.org/abs/0802.0267} {arXiv:0802.0267 [hep-th]} \BibitemShut
  {NoStop}%
\bibitem [{\citenamefont {Chen}\ and\ \citenamefont
  {Wang}(2010)}]{Chen:2009an}%
  \BibitemOpen
  \bibfield  {author} {\bibinfo {author} {\bibfnamefont {J.}~\bibnamefont
  {Chen}}\ and\ \bibinfo {author} {\bibfnamefont {Y.}~\bibnamefont {Wang}},\
  }\href {\doibase 10.1088/1674-1056/19/6/060401} {\bibfield  {journal}
  {\bibinfo  {journal} {Chin. Phys. B}\ }\textbf {\bibinfo {volume} {19}},\
  \bibinfo {pages} {060401} (\bibinfo {year} {2010})},\ \Eprint
  {http://arxiv.org/abs/0906.1318} {arXiv:0906.1318 [gr-qc]} \BibitemShut
  {NoStop}%
\bibitem [{\citenamefont {Chen}\ \emph {et~al.}(2016)\citenamefont {Chen},
  \citenamefont {Fan}, \citenamefont {Li},\ and\ \citenamefont
  {Ye}}]{Chen:2015fuf}%
  \BibitemOpen
  \bibfield  {author} {\bibinfo {author} {\bibfnamefont {B.}~\bibnamefont
  {Chen}}, \bibinfo {author} {\bibfnamefont {Z.-Y.}\ \bibnamefont {Fan}},
  \bibinfo {author} {\bibfnamefont {P.}~\bibnamefont {Li}}, \ and\ \bibinfo
  {author} {\bibfnamefont {W.}~\bibnamefont {Ye}},\ }\href {\doibase
  10.1007/JHEP01(2016)085} {\bibfield  {journal} {\bibinfo  {journal} {JHEP}\
  }\textbf {\bibinfo {volume} {01}},\ \bibinfo {pages} {085} (\bibinfo {year}
  {2016})},\ \Eprint {http://arxiv.org/abs/1511.08706} {arXiv:1511.08706
  [hep-th]} \BibitemShut {NoStop}%
\bibitem [{\citenamefont {Konoplya}\ and\ \citenamefont
  {Zinhailo}(2020)}]{Konoplya:2020bxa}%
  \BibitemOpen
  \bibfield  {author} {\bibinfo {author} {\bibfnamefont {R.~A.}\ \bibnamefont
  {Konoplya}}\ and\ \bibinfo {author} {\bibfnamefont {A.~F.}\ \bibnamefont
  {Zinhailo}},\ }\href {\doibase 10.1140/epjc/s10052-020-08639-8} {\bibfield
  {journal} {\bibinfo  {journal} {Eur. Phys. J. C}\ }\textbf {\bibinfo {volume}
  {80}},\ \bibinfo {pages} {1049} (\bibinfo {year} {2020})},\ \Eprint
  {http://arxiv.org/abs/2003.01188} {arXiv:2003.01188 [gr-qc]} \BibitemShut
  {NoStop}%
\bibitem [{\citenamefont {Yoshida}\ and\ \citenamefont
  {Soda}(2016)}]{Yoshida:2015vua}%
  \BibitemOpen
  \bibfield  {author} {\bibinfo {author} {\bibfnamefont {D.}~\bibnamefont
  {Yoshida}}\ and\ \bibinfo {author} {\bibfnamefont {J.}~\bibnamefont {Soda}},\
  }\href {\doibase 10.1103/PhysRevD.93.044024} {\bibfield  {journal} {\bibinfo
  {journal} {Phys. Rev. D}\ }\textbf {\bibinfo {volume} {93}},\ \bibinfo
  {pages} {044024} (\bibinfo {year} {2016})},\ \Eprint
  {http://arxiv.org/abs/1512.05865} {arXiv:1512.05865 [gr-qc]} \BibitemShut
  {NoStop}%
\bibitem [{\citenamefont {González}\ \emph {et~al.}(2017)\citenamefont
  {González}, \citenamefont {Konoplya},\ and\ \citenamefont
  {Vásquez}}]{Gonzalez:2017gwa}%
  \BibitemOpen
  \bibfield  {author} {\bibinfo {author} {\bibfnamefont {P.~A.}\ \bibnamefont
  {González}}, \bibinfo {author} {\bibfnamefont {R.~A.}\ \bibnamefont
  {Konoplya}}, \ and\ \bibinfo {author} {\bibfnamefont {Y.}~\bibnamefont
  {Vásquez}},\ }\href {\doibase 10.1103/PhysRevD.95.124012} {\bibfield
  {journal} {\bibinfo  {journal} {Phys. Rev. D}\ }\textbf {\bibinfo {volume}
  {95}},\ \bibinfo {pages} {124012} (\bibinfo {year} {2017})},\ \Eprint
  {http://arxiv.org/abs/1703.06215} {arXiv:1703.06215 [gr-qc]} \BibitemShut
  {NoStop}%
\bibitem [{\citenamefont {Konoplya}\ and\ \citenamefont
  {Zhidenko}(2010)}]{Konoplya:2010vz}%
  \BibitemOpen
  \bibfield  {author} {\bibinfo {author} {\bibfnamefont {R.~A.}\ \bibnamefont
  {Konoplya}}\ and\ \bibinfo {author} {\bibfnamefont {A.}~\bibnamefont
  {Zhidenko}},\ }\href {\doibase 10.1103/PhysRevD.82.084003} {\bibfield
  {journal} {\bibinfo  {journal} {Phys. Rev. D}\ }\textbf {\bibinfo {volume}
  {82}},\ \bibinfo {pages} {084003} (\bibinfo {year} {2010})},\ \Eprint
  {http://arxiv.org/abs/1004.3772} {arXiv:1004.3772 [hep-th]} \BibitemShut
  {NoStop}%
\bibitem [{\citenamefont {Berti}\ \emph {et~al.}(2003)\citenamefont {Berti},
  \citenamefont {Kokkotas},\ and\ \citenamefont
  {Papantonopoulos}}]{Berti:2003yr}%
  \BibitemOpen
  \bibfield  {author} {\bibinfo {author} {\bibfnamefont {E.}~\bibnamefont
  {Berti}}, \bibinfo {author} {\bibfnamefont {K.~D.}\ \bibnamefont {Kokkotas}},
  \ and\ \bibinfo {author} {\bibfnamefont {E.}~\bibnamefont
  {Papantonopoulos}},\ }\href {\doibase 10.1103/PhysRevD.68.064020} {\bibfield
  {journal} {\bibinfo  {journal} {Phys. Rev. D}\ }\textbf {\bibinfo {volume}
  {68}},\ \bibinfo {pages} {064020} (\bibinfo {year} {2003})},\ \Eprint
  {http://arxiv.org/abs/gr-qc/0306106} {arXiv:gr-qc/0306106} \BibitemShut
  {NoStop}%
\bibitem [{\citenamefont {Zinhailo}(2024{\natexlab{a}})}]{Zinhailo:2024jzt}%
  \BibitemOpen
  \bibfield  {author} {\bibinfo {author} {\bibfnamefont {A.~F.}\ \bibnamefont
  {Zinhailo}},\ }\href {\doibase 10.1016/j.physletb.2024.138682} {\bibfield
  {journal} {\bibinfo  {journal} {Phys. Lett. B}\ }\textbf {\bibinfo {volume}
  {853}},\ \bibinfo {pages} {138682} (\bibinfo {year} {2024}{\natexlab{a}})},\
  \Eprint {http://arxiv.org/abs/2403.06867} {arXiv:2403.06867 [gr-qc]}
  \BibitemShut {NoStop}%
\bibitem [{\citenamefont {Kanti}\ and\ \citenamefont
  {Konoplya}(2006)}]{Kanti:2005xa}%
  \BibitemOpen
  \bibfield  {author} {\bibinfo {author} {\bibfnamefont {P.}~\bibnamefont
  {Kanti}}\ and\ \bibinfo {author} {\bibfnamefont {R.~A.}\ \bibnamefont
  {Konoplya}},\ }\href {\doibase 10.1103/PhysRevD.73.044002} {\bibfield
  {journal} {\bibinfo  {journal} {Phys. Rev. D}\ }\textbf {\bibinfo {volume}
  {73}},\ \bibinfo {pages} {044002} (\bibinfo {year} {2006})},\ \Eprint
  {http://arxiv.org/abs/hep-th/0512257} {arXiv:hep-th/0512257} \BibitemShut
  {NoStop}%
\bibitem [{\citenamefont {Kanti}(2004)}]{Kanti:2004nr}%
  \BibitemOpen
  \bibfield  {author} {\bibinfo {author} {\bibfnamefont {P.}~\bibnamefont
  {Kanti}},\ }\href {\doibase 10.1142/S0217751X04018324} {\bibfield  {journal}
  {\bibinfo  {journal} {Int. J. Mod. Phys. A}\ }\textbf {\bibinfo {volume}
  {19}},\ \bibinfo {pages} {4899} (\bibinfo {year} {2004})},\ \Eprint
  {http://arxiv.org/abs/hep-ph/0402168} {arXiv:hep-ph/0402168} \BibitemShut
  {NoStop}%
\bibitem [{\citenamefont {Dubinsky}(2024)}]{Dubinsky:2024jqi}%
  \BibitemOpen
  \bibfield  {author} {\bibinfo {author} {\bibfnamefont {A.}~\bibnamefont
  {Dubinsky}},\ }\href {\doibase 10.1209/0295-5075/ad51a3} {\bibfield
  {journal} {\bibinfo  {journal} {EPL}\ }\textbf {\bibinfo {volume} {147}},\
  \bibinfo {pages} {19003} (\bibinfo {year} {2024})},\ \Eprint
  {http://arxiv.org/abs/2403.01883} {arXiv:2403.01883 [gr-qc]} \BibitemShut
  {NoStop}%
\bibitem [{\citenamefont {Dotti}\ and\ \citenamefont
  {Gleiser}(2005)}]{Dotti:2005sq}%
  \BibitemOpen
  \bibfield  {author} {\bibinfo {author} {\bibfnamefont {G.}~\bibnamefont
  {Dotti}}\ and\ \bibinfo {author} {\bibfnamefont {R.~J.}\ \bibnamefont
  {Gleiser}},\ }\href {\doibase 10.1103/PhysRevD.72.044018} {\bibfield
  {journal} {\bibinfo  {journal} {Phys. Rev. D}\ }\textbf {\bibinfo {volume}
  {72}},\ \bibinfo {pages} {044018} (\bibinfo {year} {2005})},\ \Eprint
  {http://arxiv.org/abs/gr-qc/0503117} {arXiv:gr-qc/0503117} \BibitemShut
  {NoStop}%
\bibitem [{\citenamefont {Gleiser}\ and\ \citenamefont
  {Dotti}(2005)}]{Gleiser:2005ra}%
  \BibitemOpen
  \bibfield  {author} {\bibinfo {author} {\bibfnamefont {R.~J.}\ \bibnamefont
  {Gleiser}}\ and\ \bibinfo {author} {\bibfnamefont {G.}~\bibnamefont
  {Dotti}},\ }\href {\doibase 10.1103/PhysRevD.72.124002} {\bibfield  {journal}
  {\bibinfo  {journal} {Phys. Rev. D}\ }\textbf {\bibinfo {volume} {72}},\
  \bibinfo {pages} {124002} (\bibinfo {year} {2005})},\ \Eprint
  {http://arxiv.org/abs/gr-qc/0510069} {arXiv:gr-qc/0510069} \BibitemShut
  {NoStop}%
\bibitem [{\citenamefont {Cuyubamba}\ \emph {et~al.}(2016)\citenamefont
  {Cuyubamba}, \citenamefont {Konoplya},\ and\ \citenamefont
  {Zhidenko}}]{Cuyubamba:2016cug}%
  \BibitemOpen
  \bibfield  {author} {\bibinfo {author} {\bibfnamefont {M.~A.}\ \bibnamefont
  {Cuyubamba}}, \bibinfo {author} {\bibfnamefont {R.~A.}\ \bibnamefont
  {Konoplya}}, \ and\ \bibinfo {author} {\bibfnamefont {A.}~\bibnamefont
  {Zhidenko}},\ }\href {\doibase 10.1103/PhysRevD.93.104053} {\bibfield
  {journal} {\bibinfo  {journal} {Phys. Rev. D}\ }\textbf {\bibinfo {volume}
  {93}},\ \bibinfo {pages} {104053} (\bibinfo {year} {2016})},\ \Eprint
  {http://arxiv.org/abs/1604.03604} {arXiv:1604.03604 [gr-qc]} \BibitemShut
  {NoStop}%
\bibitem [{\citenamefont {Konoplya}\ and\ \citenamefont
  {Zhidenko}(2017{\natexlab{a}})}]{Konoplya:2017ymp}%
  \BibitemOpen
  \bibfield  {author} {\bibinfo {author} {\bibfnamefont {R.~A.}\ \bibnamefont
  {Konoplya}}\ and\ \bibinfo {author} {\bibfnamefont {A.}~\bibnamefont
  {Zhidenko}},\ }\href {\doibase 10.1103/PhysRevD.95.104005} {\bibfield
  {journal} {\bibinfo  {journal} {Phys. Rev. D}\ }\textbf {\bibinfo {volume}
  {95}},\ \bibinfo {pages} {104005} (\bibinfo {year} {2017}{\natexlab{a}})},\
  \Eprint {http://arxiv.org/abs/1701.01652} {arXiv:1701.01652 [hep-th]}
  \BibitemShut {NoStop}%
\bibitem [{\citenamefont {Konoplya}\ and\ \citenamefont
  {Zhidenko}(2017{\natexlab{b}})}]{Konoplya:2017zwo}%
  \BibitemOpen
  \bibfield  {author} {\bibinfo {author} {\bibfnamefont {R.~A.}\ \bibnamefont
  {Konoplya}}\ and\ \bibinfo {author} {\bibfnamefont {A.}~\bibnamefont
  {Zhidenko}},\ }\href {\doibase 10.1007/JHEP09(2017)139} {\bibfield  {journal}
  {\bibinfo  {journal} {JHEP}\ }\textbf {\bibinfo {volume} {09}},\ \bibinfo
  {pages} {139} (\bibinfo {year} {2017}{\natexlab{b}})},\ \Eprint
  {http://arxiv.org/abs/1705.07732} {arXiv:1705.07732 [hep-th]} \BibitemShut
  {NoStop}%
\bibitem [{\citenamefont {Takahashi}\ and\ \citenamefont
  {Soda}(2010{\natexlab{a}})}]{Takahashi:2010gz}%
  \BibitemOpen
  \bibfield  {author} {\bibinfo {author} {\bibfnamefont {T.}~\bibnamefont
  {Takahashi}}\ and\ \bibinfo {author} {\bibfnamefont {J.}~\bibnamefont
  {Soda}},\ }\href {\doibase 10.1143/PTP.124.711} {\bibfield  {journal}
  {\bibinfo  {journal} {Prog. Theor. Phys.}\ }\textbf {\bibinfo {volume}
  {124}},\ \bibinfo {pages} {711} (\bibinfo {year} {2010}{\natexlab{a}})},\
  \Eprint {http://arxiv.org/abs/1008.1618} {arXiv:1008.1618 [gr-qc]}
  \BibitemShut {NoStop}%
\bibitem [{\citenamefont {Takahashi}\ and\ \citenamefont
  {Soda}(2012)}]{Takahashi:2011du}%
  \BibitemOpen
  \bibfield  {author} {\bibinfo {author} {\bibfnamefont {T.}~\bibnamefont
  {Takahashi}}\ and\ \bibinfo {author} {\bibfnamefont {J.}~\bibnamefont
  {Soda}},\ }\href {\doibase 10.1088/0264-9381/29/3/035008} {\bibfield
  {journal} {\bibinfo  {journal} {Class. Quant. Grav.}\ }\textbf {\bibinfo
  {volume} {29}},\ \bibinfo {pages} {035008} (\bibinfo {year} {2012})},\
  \Eprint {http://arxiv.org/abs/1108.5041} {arXiv:1108.5041 [hep-th]}
  \BibitemShut {NoStop}%
\bibitem [{\citenamefont {Takahashi}(2011)}]{Takahashi:2011qda}%
  \BibitemOpen
  \bibfield  {author} {\bibinfo {author} {\bibfnamefont {T.}~\bibnamefont
  {Takahashi}},\ }\href {\doibase 10.1143/PTP.125.1289} {\bibfield  {journal}
  {\bibinfo  {journal} {Prog. Theor. Phys.}\ }\textbf {\bibinfo {volume}
  {125}},\ \bibinfo {pages} {1289} (\bibinfo {year} {2011})},\ \Eprint
  {http://arxiv.org/abs/1102.1785} {arXiv:1102.1785 [gr-qc]} \BibitemShut
  {NoStop}%
\bibitem [{\citenamefont {Konoplya}\ and\ \citenamefont
  {Zhidenko}(2017{\natexlab{c}})}]{Konoplya:2017lhs}%
  \BibitemOpen
  \bibfield  {author} {\bibinfo {author} {\bibfnamefont {R.~A.}\ \bibnamefont
  {Konoplya}}\ and\ \bibinfo {author} {\bibfnamefont {A.}~\bibnamefont
  {Zhidenko}},\ }\href {\doibase 10.1088/1475-7516/2017/05/050} {\bibfield
  {journal} {\bibinfo  {journal} {JCAP}\ }\textbf {\bibinfo {volume} {05}},\
  \bibinfo {pages} {050} (\bibinfo {year} {2017}{\natexlab{c}})},\ \Eprint
  {http://arxiv.org/abs/1705.01656} {arXiv:1705.01656 [hep-th]} \BibitemShut
  {NoStop}%
\bibitem [{\citenamefont {Konoplya}\ and\ \citenamefont
  {Zhidenko}(2024{\natexlab{a}})}]{Konoplya:2022pbc}%
  \BibitemOpen
  \bibfield  {author} {\bibinfo {author} {\bibfnamefont {R.~A.}\ \bibnamefont
  {Konoplya}}\ and\ \bibinfo {author} {\bibfnamefont {A.}~\bibnamefont
  {Zhidenko}},\ }\href {\doibase 10.1016/j.jheap.2024.10.015} {\bibfield
  {journal} {\bibinfo  {journal} {JHEAp}\ }\textbf {\bibinfo {volume} {44}},\
  \bibinfo {pages} {419} (\bibinfo {year} {2024}{\natexlab{a}})},\ \Eprint
  {http://arxiv.org/abs/2209.00679} {arXiv:2209.00679 [gr-qc]} \BibitemShut
  {NoStop}%
\bibitem [{\citenamefont {Konoplya}(2023{\natexlab{a}})}]{Konoplya:2023hqb}%
  \BibitemOpen
  \bibfield  {author} {\bibinfo {author} {\bibfnamefont {R.~A.}\ \bibnamefont
  {Konoplya}},\ }\href {\doibase 10.1142/S0218271823420142} {\bibfield
  {journal} {\bibinfo  {journal} {Int. J. Mod. Phys. D}\ }\textbf {\bibinfo
  {volume} {32}},\ \bibinfo {pages} {2342014} (\bibinfo {year}
  {2023}{\natexlab{a}})},\ \Eprint {http://arxiv.org/abs/2312.16249}
  {arXiv:2312.16249 [gr-qc]} \BibitemShut {NoStop}%
\bibitem [{\citenamefont {Konoplya}\ \emph
  {et~al.}(2023{\natexlab{a}})\citenamefont {Konoplya}, \citenamefont
  {Ovchinnikov},\ and\ \citenamefont {Ahmedov}}]{Konoplya:2023ahd}%
  \BibitemOpen
  \bibfield  {author} {\bibinfo {author} {\bibfnamefont {R.~A.}\ \bibnamefont
  {Konoplya}}, \bibinfo {author} {\bibfnamefont {D.}~\bibnamefont
  {Ovchinnikov}}, \ and\ \bibinfo {author} {\bibfnamefont {B.}~\bibnamefont
  {Ahmedov}},\ }\href {\doibase 10.1103/PhysRevD.108.104054} {\bibfield
  {journal} {\bibinfo  {journal} {Phys. Rev. D}\ }\textbf {\bibinfo {volume}
  {108}},\ \bibinfo {pages} {104054} (\bibinfo {year} {2023}{\natexlab{a}})},\
  \Eprint {http://arxiv.org/abs/2307.10801} {arXiv:2307.10801 [gr-qc]}
  \BibitemShut {NoStop}%
\bibitem [{\citenamefont {Konoplya}(2023{\natexlab{b}})}]{Konoplya:2023ppx}%
  \BibitemOpen
  \bibfield  {author} {\bibinfo {author} {\bibfnamefont {R.~A.}\ \bibnamefont
  {Konoplya}},\ }\href {\doibase 10.1088/1475-7516/2023/07/001} {\bibfield
  {journal} {\bibinfo  {journal} {JCAP}\ }\textbf {\bibinfo {volume} {07}},\
  \bibinfo {pages} {001} (\bibinfo {year} {2023}{\natexlab{b}})},\ \Eprint
  {http://arxiv.org/abs/2305.09187} {arXiv:2305.09187 [gr-qc]} \BibitemShut
  {NoStop}%
\bibitem [{\citenamefont {Konoplya}(2023{\natexlab{c}})}]{Konoplya:2022iyn}%
  \BibitemOpen
  \bibfield  {author} {\bibinfo {author} {\bibfnamefont {R.~A.}\ \bibnamefont
  {Konoplya}},\ }\href {\doibase 10.1103/PhysRevD.107.064039} {\bibfield
  {journal} {\bibinfo  {journal} {Phys. Rev. D}\ }\textbf {\bibinfo {volume}
  {107}},\ \bibinfo {pages} {064039} (\bibinfo {year} {2023}{\natexlab{c}})},\
  \Eprint {http://arxiv.org/abs/2210.14506} {arXiv:2210.14506 [gr-qc]}
  \BibitemShut {NoStop}%
\bibitem [{\citenamefont {Konoplya}\ and\ \citenamefont
  {Stashko}(2025{\natexlab{a}})}]{Konoplya:2025hgp}%
  \BibitemOpen
  \bibfield  {author} {\bibinfo {author} {\bibfnamefont {R.~A.}\ \bibnamefont
  {Konoplya}}\ and\ \bibinfo {author} {\bibfnamefont {O.~S.}\ \bibnamefont
  {Stashko}},\ }\href {\doibase 10.1103/PhysRevD.111.084031} {\bibfield
  {journal} {\bibinfo  {journal} {Phys. Rev. D}\ }\textbf {\bibinfo {volume}
  {111}},\ \bibinfo {pages} {084031} (\bibinfo {year} {2025}{\natexlab{a}})},\
  \Eprint {http://arxiv.org/abs/2502.05689} {arXiv:2502.05689 [gr-qc]}
  \BibitemShut {NoStop}%
\bibitem [{\citenamefont {Konoplya}\ and\ \citenamefont
  {Stashko}(2025{\natexlab{b}})}]{Konoplya:2024lch}%
  \BibitemOpen
  \bibfield  {author} {\bibinfo {author} {\bibfnamefont {R.~A.}\ \bibnamefont
  {Konoplya}}\ and\ \bibinfo {author} {\bibfnamefont {O.~S.}\ \bibnamefont
  {Stashko}},\ }\href {\doibase 10.1103/PhysRevD.111.104055} {\bibfield
  {journal} {\bibinfo  {journal} {Phys. Rev. D}\ }\textbf {\bibinfo {volume}
  {111}},\ \bibinfo {pages} {104055} (\bibinfo {year} {2025}{\natexlab{b}})},\
  \Eprint {http://arxiv.org/abs/2408.02578} {arXiv:2408.02578 [gr-qc]}
  \BibitemShut {NoStop}%
\bibitem [{\citenamefont {Zhang}\ \emph {et~al.}(2024)\citenamefont {Zhang},
  \citenamefont {Gong}, \citenamefont {Fu}, \citenamefont {Wu},\ and\
  \citenamefont {Pan}}]{Zhang:2024nny}%
  \BibitemOpen
  \bibfield  {author} {\bibinfo {author} {\bibfnamefont {D.}~\bibnamefont
  {Zhang}}, \bibinfo {author} {\bibfnamefont {H.}~\bibnamefont {Gong}},
  \bibinfo {author} {\bibfnamefont {G.}~\bibnamefont {Fu}}, \bibinfo {author}
  {\bibfnamefont {J.-P.}\ \bibnamefont {Wu}}, \ and\ \bibinfo {author}
  {\bibfnamefont {Q.}~\bibnamefont {Pan}},\ }\href {\doibase
  10.1140/epjc/s10052-024-12928-x} {\bibfield  {journal} {\bibinfo  {journal}
  {Eur. Phys. J. C}\ }\textbf {\bibinfo {volume} {84}},\ \bibinfo {pages} {564}
  (\bibinfo {year} {2024})},\ \Eprint {http://arxiv.org/abs/2402.15085}
  {arXiv:2402.15085 [gr-qc]} \BibitemShut {NoStop}%
\bibitem [{\citenamefont {Bolokhov}(2024{\natexlab{a}})}]{Bolokhov:2023bwm}%
  \BibitemOpen
  \bibfield  {author} {\bibinfo {author} {\bibfnamefont {S.~V.}\ \bibnamefont
  {Bolokhov}},\ }\href {\doibase 10.1103/PhysRevD.110.024010} {\bibfield
  {journal} {\bibinfo  {journal} {Phys. Rev. D}\ }\textbf {\bibinfo {volume}
  {110}},\ \bibinfo {pages} {024010} (\bibinfo {year} {2024}{\natexlab{a}})},\
  \Eprint {http://arxiv.org/abs/2311.05503} {arXiv:2311.05503 [gr-qc]}
  \BibitemShut {NoStop}%
\bibitem [{\citenamefont {Bolokhov}(2024{\natexlab{b}})}]{Bolokhov:2023ruj}%
  \BibitemOpen
  \bibfield  {author} {\bibinfo {author} {\bibfnamefont {S.~V.}\ \bibnamefont
  {Bolokhov}},\ }\href {\doibase 10.1103/PhysRevD.109.064017} {\bibfield
  {journal} {\bibinfo  {journal} {Phys. Rev. D}\ }\textbf {\bibinfo {volume}
  {109}},\ \bibinfo {pages} {064017} (\bibinfo {year}
  {2024}{\natexlab{b}})}\BibitemShut {NoStop}%
\bibitem [{\citenamefont
  {Lütfüoğlu}(2025{\natexlab{a}})}]{Lutfuoglu:2025ljm}%
  \BibitemOpen
  \bibfield  {author} {\bibinfo {author} {\bibfnamefont {B.~C.}\ \bibnamefont
  {Lütfüoğlu}},\ }\href {\doibase 10.1140/epjc/s10052-025-14380-x}
  {\bibfield  {journal} {\bibinfo  {journal} {Eur. Phys. J. C}\ }\textbf
  {\bibinfo {volume} {85}},\ \bibinfo {pages} {630} (\bibinfo {year}
  {2025}{\natexlab{a}})},\ \Eprint {http://arxiv.org/abs/2504.18482}
  {arXiv:2504.18482 [gr-qc]} \BibitemShut {NoStop}%
\bibitem [{\citenamefont {Stashko}(2024)}]{Stashko:2024wuq}%
  \BibitemOpen
  \bibfield  {author} {\bibinfo {author} {\bibfnamefont {O.}~\bibnamefont
  {Stashko}},\ }\href {\doibase 10.1103/PhysRevD.110.084016} {\bibfield
  {journal} {\bibinfo  {journal} {Phys. Rev. D}\ }\textbf {\bibinfo {volume}
  {110}},\ \bibinfo {pages} {084016} (\bibinfo {year} {2024})},\ \Eprint
  {http://arxiv.org/abs/2407.07892} {arXiv:2407.07892 [gr-qc]} \BibitemShut
  {NoStop}%
\bibitem [{\citenamefont {Zinhailo}(2024{\natexlab{b}})}]{Zinhailo:2024kbq}%
  \BibitemOpen
  \bibfield  {author} {\bibinfo {author} {\bibfnamefont {A.~F.}\ \bibnamefont
  {Zinhailo}},\ }\href {\doibase 10.13140/RG.2.2.26785.01124} {\  (\bibinfo
  {year} {2024}{\natexlab{b}}),\ 10.13140/RG.2.2.26785.01124}\BibitemShut
  {NoStop}%
\bibitem [{\citenamefont {Moura}\ and\ \citenamefont
  {Rodrigues}(2023)}]{Moura:2022gqm}%
  \BibitemOpen
  \bibfield  {author} {\bibinfo {author} {\bibfnamefont {F.}~\bibnamefont
  {Moura}}\ and\ \bibinfo {author} {\bibfnamefont {J.}~\bibnamefont
  {Rodrigues}},\ }\href {\doibase 10.1016/j.nuclphysb.2023.116255} {\bibfield
  {journal} {\bibinfo  {journal} {Nucl. Phys. B}\ }\textbf {\bibinfo {volume}
  {993}},\ \bibinfo {pages} {116255} (\bibinfo {year} {2023})},\ \Eprint
  {http://arxiv.org/abs/2206.11377} {arXiv:2206.11377 [hep-th]} \BibitemShut
  {NoStop}%
\bibitem [{\citenamefont {Cao}\ \emph {et~al.}(2024)\citenamefont {Cao},
  \citenamefont {Wu},\ and\ \citenamefont {Zhou}}]{Cao:2024sot}%
  \BibitemOpen
  \bibfield  {author} {\bibinfo {author} {\bibfnamefont {L.-M.}\ \bibnamefont
  {Cao}}, \bibinfo {author} {\bibfnamefont {L.-B.}\ \bibnamefont {Wu}}, \ and\
  \bibinfo {author} {\bibfnamefont {Y.-S.}\ \bibnamefont {Zhou}},\ }\href@noop
  {} {\  (\bibinfo {year} {2024})},\ \Eprint {http://arxiv.org/abs/2412.21092}
  {arXiv:2412.21092 [gr-qc]} \BibitemShut {NoStop}%
\bibitem [{\citenamefont {Leaver}(1985)}]{Leaver:1985ax}%
  \BibitemOpen
  \bibfield  {author} {\bibinfo {author} {\bibfnamefont {E.~W.}\ \bibnamefont
  {Leaver}},\ }\href {\doibase 10.1098/rspa.1985.0119} {\bibfield  {journal}
  {\bibinfo  {journal} {Proc. Roy. Soc. Lond. A}\ }\textbf {\bibinfo {volume}
  {402}},\ \bibinfo {pages} {285} (\bibinfo {year} {1985})}\BibitemShut
  {NoStop}%
\bibitem [{\citenamefont {Konoplya}\ \emph
  {et~al.}(2023{\natexlab{b}})\citenamefont {Konoplya}, \citenamefont
  {Stuchlik}, \citenamefont {Zhidenko},\ and\ \citenamefont
  {Zinhailo}}]{Konoplya:2023aph}%
  \BibitemOpen
  \bibfield  {author} {\bibinfo {author} {\bibfnamefont {R.~A.}\ \bibnamefont
  {Konoplya}}, \bibinfo {author} {\bibfnamefont {Z.}~\bibnamefont {Stuchlik}},
  \bibinfo {author} {\bibfnamefont {A.}~\bibnamefont {Zhidenko}}, \ and\
  \bibinfo {author} {\bibfnamefont {A.~F.}\ \bibnamefont {Zinhailo}},\ }\href
  {\doibase 10.1103/PhysRevD.107.104050} {\bibfield  {journal} {\bibinfo
  {journal} {Phys. Rev. D}\ }\textbf {\bibinfo {volume} {107}},\ \bibinfo
  {pages} {104050} (\bibinfo {year} {2023}{\natexlab{b}})},\ \Eprint
  {http://arxiv.org/abs/2303.01987} {arXiv:2303.01987 [gr-qc]} \BibitemShut
  {NoStop}%
\bibitem [{\citenamefont {Tangherlini}(1963)}]{Tangherlini:1963bw}%
  \BibitemOpen
  \bibfield  {author} {\bibinfo {author} {\bibfnamefont {F.~R.}\ \bibnamefont
  {Tangherlini}},\ }\href {\doibase 10.1007/BF02784569} {\bibfield  {journal}
  {\bibinfo  {journal} {Nuovo Cim.}\ }\textbf {\bibinfo {volume} {27}},\
  \bibinfo {pages} {636} (\bibinfo {year} {1963})}\BibitemShut {NoStop}%
\bibitem [{\citenamefont {Takahashi}\ and\ \citenamefont
  {Soda}(2010{\natexlab{b}})}]{Takahashi:2010ye}%
  \BibitemOpen
  \bibfield  {author} {\bibinfo {author} {\bibfnamefont {T.}~\bibnamefont
  {Takahashi}}\ and\ \bibinfo {author} {\bibfnamefont {J.}~\bibnamefont
  {Soda}},\ }\href {\doibase 10.1143/PTP.124.911} {\bibfield  {journal}
  {\bibinfo  {journal} {Prog. Theor. Phys.}\ }\textbf {\bibinfo {volume}
  {124}},\ \bibinfo {pages} {911} (\bibinfo {year} {2010}{\natexlab{b}})},\
  \Eprint {http://arxiv.org/abs/1008.1385} {arXiv:1008.1385 [gr-qc]}
  \BibitemShut {NoStop}%
\bibitem [{\citenamefont {Reall}\ \emph {et~al.}(2014)\citenamefont {Reall},
  \citenamefont {Tanahashi},\ and\ \citenamefont {Way}}]{Reall:2014pwa}%
  \BibitemOpen
  \bibfield  {author} {\bibinfo {author} {\bibfnamefont {H.}~\bibnamefont
  {Reall}}, \bibinfo {author} {\bibfnamefont {N.}~\bibnamefont {Tanahashi}}, \
  and\ \bibinfo {author} {\bibfnamefont {B.}~\bibnamefont {Way}},\ }\href
  {\doibase 10.1088/0264-9381/31/20/205005} {\bibfield  {journal} {\bibinfo
  {journal} {Class. Quant. Grav.}\ }\textbf {\bibinfo {volume} {31}},\ \bibinfo
  {pages} {205005} (\bibinfo {year} {2014})},\ \Eprint
  {http://arxiv.org/abs/1406.3379} {arXiv:1406.3379 [hep-th]} \BibitemShut
  {NoStop}%
\bibitem [{\citenamefont {Rezzolla}\ and\ \citenamefont
  {Zhidenko}(2014)}]{Rezzolla:2014mua}%
  \BibitemOpen
  \bibfield  {author} {\bibinfo {author} {\bibfnamefont {L.}~\bibnamefont
  {Rezzolla}}\ and\ \bibinfo {author} {\bibfnamefont {A.}~\bibnamefont
  {Zhidenko}},\ }\href {\doibase 10.1103/PhysRevD.90.084009} {\bibfield
  {journal} {\bibinfo  {journal} {Phys. Rev. D}\ }\textbf {\bibinfo {volume}
  {90}},\ \bibinfo {pages} {084009} (\bibinfo {year} {2014})},\ \Eprint
  {http://arxiv.org/abs/1407.3086} {arXiv:1407.3086 [gr-qc]} \BibitemShut
  {NoStop}%
\bibitem [{\citenamefont {Konoplya}\ \emph {et~al.}(2016)\citenamefont
  {Konoplya}, \citenamefont {Rezzolla},\ and\ \citenamefont
  {Zhidenko}}]{Konoplya:2016jvv}%
  \BibitemOpen
  \bibfield  {author} {\bibinfo {author} {\bibfnamefont {R.}~\bibnamefont
  {Konoplya}}, \bibinfo {author} {\bibfnamefont {L.}~\bibnamefont {Rezzolla}},
  \ and\ \bibinfo {author} {\bibfnamefont {A.}~\bibnamefont {Zhidenko}},\
  }\href {\doibase 10.1103/PhysRevD.93.064015} {\bibfield  {journal} {\bibinfo
  {journal} {Phys. Rev. D}\ }\textbf {\bibinfo {volume} {93}},\ \bibinfo
  {pages} {064015} (\bibinfo {year} {2016})},\ \Eprint
  {http://arxiv.org/abs/1602.02378} {arXiv:1602.02378 [gr-qc]} \BibitemShut
  {NoStop}%
\bibitem [{\citenamefont {Bronnikov}\ \emph {et~al.}(2021)\citenamefont
  {Bronnikov}, \citenamefont {Konoplya},\ and\ \citenamefont
  {Pappas}}]{Bronnikov:2021liv}%
  \BibitemOpen
  \bibfield  {author} {\bibinfo {author} {\bibfnamefont {K.~A.}\ \bibnamefont
  {Bronnikov}}, \bibinfo {author} {\bibfnamefont {R.~A.}\ \bibnamefont
  {Konoplya}}, \ and\ \bibinfo {author} {\bibfnamefont {T.~D.}\ \bibnamefont
  {Pappas}},\ }\href {\doibase 10.1103/PhysRevD.103.124062} {\bibfield
  {journal} {\bibinfo  {journal} {Phys. Rev. D}\ }\textbf {\bibinfo {volume}
  {103}},\ \bibinfo {pages} {124062} (\bibinfo {year} {2021})},\ \Eprint
  {http://arxiv.org/abs/2102.10679} {arXiv:2102.10679 [gr-qc]} \BibitemShut
  {NoStop}%
\bibitem [{\citenamefont {Younsi}\ \emph {et~al.}(2016)\citenamefont {Younsi},
  \citenamefont {Zhidenko}, \citenamefont {Rezzolla}, \citenamefont
  {Konoplya},\ and\ \citenamefont {Mizuno}}]{Younsi:2016azx}%
  \BibitemOpen
  \bibfield  {author} {\bibinfo {author} {\bibfnamefont {Z.}~\bibnamefont
  {Younsi}}, \bibinfo {author} {\bibfnamefont {A.}~\bibnamefont {Zhidenko}},
  \bibinfo {author} {\bibfnamefont {L.}~\bibnamefont {Rezzolla}}, \bibinfo
  {author} {\bibfnamefont {R.}~\bibnamefont {Konoplya}}, \ and\ \bibinfo
  {author} {\bibfnamefont {Y.}~\bibnamefont {Mizuno}},\ }\href {\doibase
  10.1103/PhysRevD.94.084025} {\bibfield  {journal} {\bibinfo  {journal} {Phys.
  Rev. D}\ }\textbf {\bibinfo {volume} {94}},\ \bibinfo {pages} {084025}
  (\bibinfo {year} {2016})},\ \Eprint {http://arxiv.org/abs/1607.05767}
  {arXiv:1607.05767 [gr-qc]} \BibitemShut {NoStop}%
\bibitem [{\citenamefont {Kokkotas}\ \emph
  {et~al.}(2017{\natexlab{a}})\citenamefont {Kokkotas}, \citenamefont
  {Konoplya},\ and\ \citenamefont {Zhidenko}}]{Kokkotas:2017zwt}%
  \BibitemOpen
  \bibfield  {author} {\bibinfo {author} {\bibfnamefont {K.}~\bibnamefont
  {Kokkotas}}, \bibinfo {author} {\bibfnamefont {R.~A.}\ \bibnamefont
  {Konoplya}}, \ and\ \bibinfo {author} {\bibfnamefont {A.}~\bibnamefont
  {Zhidenko}},\ }\href {\doibase 10.1103/PhysRevD.96.064007} {\bibfield
  {journal} {\bibinfo  {journal} {Phys. Rev. D}\ }\textbf {\bibinfo {volume}
  {96}},\ \bibinfo {pages} {064007} (\bibinfo {year} {2017}{\natexlab{a}})},\
  \Eprint {http://arxiv.org/abs/1705.09875} {arXiv:1705.09875 [gr-qc]}
  \BibitemShut {NoStop}%
\bibitem [{\citenamefont {Kokkotas}\ \emph
  {et~al.}(2017{\natexlab{b}})\citenamefont {Kokkotas}, \citenamefont
  {Konoplya},\ and\ \citenamefont {Zhidenko}}]{Kokkotas:2017ymc}%
  \BibitemOpen
  \bibfield  {author} {\bibinfo {author} {\bibfnamefont {K.~D.}\ \bibnamefont
  {Kokkotas}}, \bibinfo {author} {\bibfnamefont {R.~A.}\ \bibnamefont
  {Konoplya}}, \ and\ \bibinfo {author} {\bibfnamefont {A.}~\bibnamefont
  {Zhidenko}},\ }\href {\doibase 10.1103/PhysRevD.96.064004} {\bibfield
  {journal} {\bibinfo  {journal} {Phys. Rev. D}\ }\textbf {\bibinfo {volume}
  {96}},\ \bibinfo {pages} {064004} (\bibinfo {year} {2017}{\natexlab{b}})},\
  \Eprint {http://arxiv.org/abs/1706.07460} {arXiv:1706.07460 [gr-qc]}
  \BibitemShut {NoStop}%
\bibitem [{\citenamefont {Konoplya}\ \emph {et~al.}(2018)\citenamefont
  {Konoplya}, \citenamefont {Stuchlík},\ and\ \citenamefont
  {Zhidenko}}]{Konoplya:2018arm}%
  \BibitemOpen
  \bibfield  {author} {\bibinfo {author} {\bibfnamefont {R.~A.}\ \bibnamefont
  {Konoplya}}, \bibinfo {author} {\bibfnamefont {Z.}~\bibnamefont {Stuchlík}},
  \ and\ \bibinfo {author} {\bibfnamefont {A.}~\bibnamefont {Zhidenko}},\
  }\href {\doibase 10.1103/PhysRevD.97.084044} {\bibfield  {journal} {\bibinfo
  {journal} {Phys. Rev. D}\ }\textbf {\bibinfo {volume} {97}},\ \bibinfo
  {pages} {084044} (\bibinfo {year} {2018})},\ \Eprint
  {http://arxiv.org/abs/1801.07195} {arXiv:1801.07195 [gr-qc]} \BibitemShut
  {NoStop}%
\bibitem [{\citenamefont {Zinhailo}(2019)}]{Zinhailo:2019rwd}%
  \BibitemOpen
  \bibfield  {author} {\bibinfo {author} {\bibfnamefont {A.~F.}\ \bibnamefont
  {Zinhailo}},\ }\href {\doibase 10.1140/epjc/s10052-019-7425-9} {\bibfield
  {journal} {\bibinfo  {journal} {Eur. Phys. J. C}\ }\textbf {\bibinfo {volume}
  {79}},\ \bibinfo {pages} {912} (\bibinfo {year} {2019})},\ \Eprint
  {http://arxiv.org/abs/1909.12664} {arXiv:1909.12664 [gr-qc]} \BibitemShut
  {NoStop}%
\bibitem [{\citenamefont {Konoplya}\ and\ \citenamefont
  {Zinhailo}(2019)}]{Konoplya:2019ppy}%
  \BibitemOpen
  \bibfield  {author} {\bibinfo {author} {\bibfnamefont {R.~A.}\ \bibnamefont
  {Konoplya}}\ and\ \bibinfo {author} {\bibfnamefont {A.~F.}\ \bibnamefont
  {Zinhailo}},\ }\href {\doibase 10.1103/PhysRevD.99.104060} {\bibfield
  {journal} {\bibinfo  {journal} {Phys. Rev. D}\ }\textbf {\bibinfo {volume}
  {99}},\ \bibinfo {pages} {104060} (\bibinfo {year} {2019})},\ \Eprint
  {http://arxiv.org/abs/1904.05341} {arXiv:1904.05341 [gr-qc]} \BibitemShut
  {NoStop}%
\bibitem [{\citenamefont {Nampalliwar}\ \emph {et~al.}(2018)\citenamefont
  {Nampalliwar}, \citenamefont {Bambi}, \citenamefont {Kokkotas},\ and\
  \citenamefont {Konoplya}}]{Nampalliwar:2018iru}%
  \BibitemOpen
  \bibfield  {author} {\bibinfo {author} {\bibfnamefont {S.}~\bibnamefont
  {Nampalliwar}}, \bibinfo {author} {\bibfnamefont {C.}~\bibnamefont {Bambi}},
  \bibinfo {author} {\bibfnamefont {K.}~\bibnamefont {Kokkotas}}, \ and\
  \bibinfo {author} {\bibfnamefont {R.}~\bibnamefont {Konoplya}},\ }\href
  {\doibase 10.1016/j.physletb.2018.04.053} {\bibfield  {journal} {\bibinfo
  {journal} {Phys. Lett. B}\ }\textbf {\bibinfo {volume} {781}},\ \bibinfo
  {pages} {626} (\bibinfo {year} {2018})},\ \Eprint
  {http://arxiv.org/abs/1803.10819} {arXiv:1803.10819 [gr-qc]} \BibitemShut
  {NoStop}%
\bibitem [{\citenamefont {Konoplya}\ and\ \citenamefont
  {Zhidenko}(2023)}]{Konoplya:2023owh}%
  \BibitemOpen
  \bibfield  {author} {\bibinfo {author} {\bibfnamefont {R.~A.}\ \bibnamefont
  {Konoplya}}\ and\ \bibinfo {author} {\bibfnamefont {A.}~\bibnamefont
  {Zhidenko}},\ }\href {\doibase 10.1088/1475-7516/2023/08/008} {\bibfield
  {journal} {\bibinfo  {journal} {JCAP}\ }\textbf {\bibinfo {volume} {08}},\
  \bibinfo {pages} {008} (\bibinfo {year} {2023})},\ \Eprint
  {http://arxiv.org/abs/2303.03130} {arXiv:2303.03130 [gr-qc]} \BibitemShut
  {NoStop}%
\bibitem [{\citenamefont {Konoplya}\ and\ \citenamefont
  {Zhidenko}(2022)}]{Konoplya:2022kld}%
  \BibitemOpen
  \bibfield  {author} {\bibinfo {author} {\bibfnamefont {R.~A.}\ \bibnamefont
  {Konoplya}}\ and\ \bibinfo {author} {\bibfnamefont {A.}~\bibnamefont
  {Zhidenko}},\ }\href {\doibase 10.1088/1475-7516/2022/11/028} {\bibfield
  {journal} {\bibinfo  {journal} {JCAP}\ }\textbf {\bibinfo {volume} {11}},\
  \bibinfo {pages} {028} (\bibinfo {year} {2022})},\ \Eprint
  {http://arxiv.org/abs/2210.04314} {arXiv:2210.04314 [gr-qc]} \BibitemShut
  {NoStop}%
\bibitem [{\citenamefont {Dubinsky}(2025{\natexlab{a}})}]{Dubinsky:2024rvf}%
  \BibitemOpen
  \bibfield  {author} {\bibinfo {author} {\bibfnamefont {A.}~\bibnamefont
  {Dubinsky}},\ }\href {\doibase 10.1016/j.physletb.2025.139251} {\bibfield
  {journal} {\bibinfo  {journal} {Phys. Lett. B}\ }\textbf {\bibinfo {volume}
  {861}},\ \bibinfo {pages} {139251} (\bibinfo {year} {2025}{\natexlab{a}})},\
  \Eprint {http://arxiv.org/abs/2409.16569} {arXiv:2409.16569 [gr-qc]}
  \BibitemShut {NoStop}%
\bibitem [{\citenamefont {Dubinsky}\ and\ \citenamefont
  {Zinhailo}(2025)}]{Dubinsky:2024nzo}%
  \BibitemOpen
  \bibfield  {author} {\bibinfo {author} {\bibfnamefont {A.}~\bibnamefont
  {Dubinsky}}\ and\ \bibinfo {author} {\bibfnamefont {A.~F.}\ \bibnamefont
  {Zinhailo}},\ }\href {\doibase 10.1209/0295-5075/adbc17} {\bibfield
  {journal} {\bibinfo  {journal} {EPL}\ }\textbf {\bibinfo {volume} {149}},\
  \bibinfo {pages} {69004} (\bibinfo {year} {2025})},\ \Eprint
  {http://arxiv.org/abs/2410.15232} {arXiv:2410.15232 [gr-qc]} \BibitemShut
  {NoStop}%
\bibitem [{\citenamefont {Kocherlakota}\ and\ \citenamefont
  {Rezzolla}(2020)}]{Kocherlakota:2020kyu}%
  \BibitemOpen
  \bibfield  {author} {\bibinfo {author} {\bibfnamefont {P.}~\bibnamefont
  {Kocherlakota}}\ and\ \bibinfo {author} {\bibfnamefont {L.}~\bibnamefont
  {Rezzolla}},\ }\href {\doibase 10.1103/PhysRevD.102.064058} {\bibfield
  {journal} {\bibinfo  {journal} {Phys. Rev. D}\ }\textbf {\bibinfo {volume}
  {102}},\ \bibinfo {pages} {064058} (\bibinfo {year} {2020})},\ \Eprint
  {http://arxiv.org/abs/2007.15593} {arXiv:2007.15593 [gr-qc]} \BibitemShut
  {NoStop}%
\bibitem [{\citenamefont {Zhang}(2024)}]{Zhang:2024rvk}%
  \BibitemOpen
  \bibfield  {author} {\bibinfo {author} {\bibfnamefont {S.-J.}\ \bibnamefont
  {Zhang}},\ }\href {\doibase 10.1103/PhysRevD.109.084066} {\bibfield
  {journal} {\bibinfo  {journal} {Phys. Rev. D}\ }\textbf {\bibinfo {volume}
  {109}},\ \bibinfo {pages} {084066} (\bibinfo {year} {2024})},\ \Eprint
  {http://arxiv.org/abs/2402.15050} {arXiv:2402.15050 [gr-qc]} \BibitemShut
  {NoStop}%
\bibitem [{\citenamefont {Cassing}\ and\ \citenamefont
  {Rezzolla}(2023)}]{Cassing:2023bpt}%
  \BibitemOpen
  \bibfield  {author} {\bibinfo {author} {\bibfnamefont {M.}~\bibnamefont
  {Cassing}}\ and\ \bibinfo {author} {\bibfnamefont {L.}~\bibnamefont
  {Rezzolla}},\ }\href {\doibase 10.1093/mnras/stad1039} {\bibfield  {journal}
  {\bibinfo  {journal} {Mon. Not. Roy. Astron. Soc.}\ }\textbf {\bibinfo
  {volume} {522}},\ \bibinfo {pages} {2415} (\bibinfo {year} {2023})},\ \Eprint
  {http://arxiv.org/abs/2302.09135} {arXiv:2302.09135 [gr-qc]} \BibitemShut
  {NoStop}%
\bibitem [{\citenamefont {Li}\ \emph {et~al.}(2021)\citenamefont {Li},
  \citenamefont {Abdujabbarov},\ and\ \citenamefont {Han}}]{Li:2021mnx}%
  \BibitemOpen
  \bibfield  {author} {\bibinfo {author} {\bibfnamefont {S.}~\bibnamefont
  {Li}}, \bibinfo {author} {\bibfnamefont {A.~A.}\ \bibnamefont
  {Abdujabbarov}}, \ and\ \bibinfo {author} {\bibfnamefont {W.-B.}\
  \bibnamefont {Han}},\ }\href {\doibase 10.1140/epjc/s10052-021-09445-6}
  {\bibfield  {journal} {\bibinfo  {journal} {Eur. Phys. J. C}\ }\textbf
  {\bibinfo {volume} {81}},\ \bibinfo {pages} {649} (\bibinfo {year} {2021})},\
  \Eprint {http://arxiv.org/abs/2103.08104} {arXiv:2103.08104 [gr-qc]}
  \BibitemShut {NoStop}%
\bibitem [{\citenamefont {Ma}\ and\ \citenamefont
  {Rezzolla}(2024)}]{Ma:2024kbu}%
  \BibitemOpen
  \bibfield  {author} {\bibinfo {author} {\bibfnamefont {Y.}~\bibnamefont
  {Ma}}\ and\ \bibinfo {author} {\bibfnamefont {L.}~\bibnamefont {Rezzolla}},\
  }\href {\doibase 10.1103/PhysRevD.110.024032} {\bibfield  {journal} {\bibinfo
   {journal} {Phys. Rev. D}\ }\textbf {\bibinfo {volume} {110}},\ \bibinfo
  {pages} {024032} (\bibinfo {year} {2024})},\ \Eprint
  {http://arxiv.org/abs/2404.06509} {arXiv:2404.06509 [gr-qc]} \BibitemShut
  {NoStop}%
\bibitem [{\citenamefont {Shashank}\ and\ \citenamefont
  {Bambi}(2022)}]{Shashank:2021giy}%
  \BibitemOpen
  \bibfield  {author} {\bibinfo {author} {\bibfnamefont {S.}~\bibnamefont
  {Shashank}}\ and\ \bibinfo {author} {\bibfnamefont {C.}~\bibnamefont
  {Bambi}},\ }\href {\doibase 10.1103/PhysRevD.105.104004} {\bibfield
  {journal} {\bibinfo  {journal} {Phys. Rev. D}\ }\textbf {\bibinfo {volume}
  {105}},\ \bibinfo {pages} {104004} (\bibinfo {year} {2022})},\ \Eprint
  {http://arxiv.org/abs/2112.05388} {arXiv:2112.05388 [gr-qc]} \BibitemShut
  {NoStop}%
\bibitem [{\citenamefont {Konoplya}\ and\ \citenamefont
  {Zhidenko}(2021)}]{Konoplya:2021slg}%
  \BibitemOpen
  \bibfield  {author} {\bibinfo {author} {\bibfnamefont {R.~A.}\ \bibnamefont
  {Konoplya}}\ and\ \bibinfo {author} {\bibfnamefont {A.}~\bibnamefont
  {Zhidenko}},\ }\href {\doibase 10.1103/PhysRevD.103.104033} {\bibfield
  {journal} {\bibinfo  {journal} {Phys. Rev. D}\ }\textbf {\bibinfo {volume}
  {103}},\ \bibinfo {pages} {104033} (\bibinfo {year} {2021})},\ \Eprint
  {http://arxiv.org/abs/2103.03855} {arXiv:2103.03855 [gr-qc]} \BibitemShut
  {NoStop}%
\bibitem [{\citenamefont {Yu}\ \emph {et~al.}(2021)\citenamefont {Yu},
  \citenamefont {Jiang}, \citenamefont {Abdikamalov}, \citenamefont
  {Ayzenberg}, \citenamefont {Bambi}, \citenamefont {Liu}, \citenamefont
  {Nampalliwar},\ and\ \citenamefont {Tripathi}}]{Yu:2021xen}%
  \BibitemOpen
  \bibfield  {author} {\bibinfo {author} {\bibfnamefont {Z.}~\bibnamefont
  {Yu}}, \bibinfo {author} {\bibfnamefont {Q.}~\bibnamefont {Jiang}}, \bibinfo
  {author} {\bibfnamefont {A.~B.}\ \bibnamefont {Abdikamalov}}, \bibinfo
  {author} {\bibfnamefont {D.}~\bibnamefont {Ayzenberg}}, \bibinfo {author}
  {\bibfnamefont {C.}~\bibnamefont {Bambi}}, \bibinfo {author} {\bibfnamefont
  {H.}~\bibnamefont {Liu}}, \bibinfo {author} {\bibfnamefont {S.}~\bibnamefont
  {Nampalliwar}}, \ and\ \bibinfo {author} {\bibfnamefont {A.}~\bibnamefont
  {Tripathi}},\ }\href {\doibase 10.1103/PhysRevD.104.084035} {\bibfield
  {journal} {\bibinfo  {journal} {Phys. Rev. D}\ }\textbf {\bibinfo {volume}
  {104}},\ \bibinfo {pages} {084035} (\bibinfo {year} {2021})},\ \Eprint
  {http://arxiv.org/abs/2106.11658} {arXiv:2106.11658 [astro-ph.HE]}
  \BibitemShut {NoStop}%
\bibitem [{\citenamefont {Toshmatov}\ and\ \citenamefont
  {Ahmedov}(2023)}]{Toshmatov:2023anz}%
  \BibitemOpen
  \bibfield  {author} {\bibinfo {author} {\bibfnamefont {B.}~\bibnamefont
  {Toshmatov}}\ and\ \bibinfo {author} {\bibfnamefont {B.}~\bibnamefont
  {Ahmedov}},\ }\href {\doibase 10.1103/PhysRevD.108.084035} {\bibfield
  {journal} {\bibinfo  {journal} {Phys. Rev. D}\ }\textbf {\bibinfo {volume}
  {108}},\ \bibinfo {pages} {084035} (\bibinfo {year} {2023})},\ \Eprint
  {http://arxiv.org/abs/2311.04602} {arXiv:2311.04602 [gr-qc]} \BibitemShut
  {NoStop}%
\bibitem [{\citenamefont {Konoplya}\ \emph
  {et~al.}(2020{\natexlab{a}})\citenamefont {Konoplya}, \citenamefont
  {Pappas},\ and\ \citenamefont {Zhidenko}}]{Konoplya:2019fpy}%
  \BibitemOpen
  \bibfield  {author} {\bibinfo {author} {\bibfnamefont {R.~A.}\ \bibnamefont
  {Konoplya}}, \bibinfo {author} {\bibfnamefont {T.}~\bibnamefont {Pappas}}, \
  and\ \bibinfo {author} {\bibfnamefont {A.}~\bibnamefont {Zhidenko}},\ }\href
  {\doibase 10.1103/PhysRevD.101.044054} {\bibfield  {journal} {\bibinfo
  {journal} {Phys. Rev. D}\ }\textbf {\bibinfo {volume} {101}},\ \bibinfo
  {pages} {044054} (\bibinfo {year} {2020}{\natexlab{a}})},\ \Eprint
  {http://arxiv.org/abs/1907.10112} {arXiv:1907.10112 [gr-qc]} \BibitemShut
  {NoStop}%
\bibitem [{\citenamefont {Nampalliwar}\ \emph {et~al.}(2020)\citenamefont
  {Nampalliwar}, \citenamefont {Xin}, \citenamefont {Srivastava}, \citenamefont
  {Abdikamalov}, \citenamefont {Ayzenberg}, \citenamefont {Bambi},
  \citenamefont {Dauser}, \citenamefont {Garcia},\ and\ \citenamefont
  {Tripathi}}]{Nampalliwar:2019iti}%
  \BibitemOpen
  \bibfield  {author} {\bibinfo {author} {\bibfnamefont {S.}~\bibnamefont
  {Nampalliwar}}, \bibinfo {author} {\bibfnamefont {S.}~\bibnamefont {Xin}},
  \bibinfo {author} {\bibfnamefont {S.}~\bibnamefont {Srivastava}}, \bibinfo
  {author} {\bibfnamefont {A.~B.}\ \bibnamefont {Abdikamalov}}, \bibinfo
  {author} {\bibfnamefont {D.}~\bibnamefont {Ayzenberg}}, \bibinfo {author}
  {\bibfnamefont {C.}~\bibnamefont {Bambi}}, \bibinfo {author} {\bibfnamefont
  {T.}~\bibnamefont {Dauser}}, \bibinfo {author} {\bibfnamefont {J.~A.}\
  \bibnamefont {Garcia}}, \ and\ \bibinfo {author} {\bibfnamefont
  {A.}~\bibnamefont {Tripathi}},\ }\href {\doibase 10.1103/PhysRevD.102.124071}
  {\bibfield  {journal} {\bibinfo  {journal} {Phys. Rev. D}\ }\textbf {\bibinfo
  {volume} {102}},\ \bibinfo {pages} {124071} (\bibinfo {year} {2020})},\
  \Eprint {http://arxiv.org/abs/1903.12119} {arXiv:1903.12119 [gr-qc]}
  \BibitemShut {NoStop}%
\bibitem [{\citenamefont {Ni}\ \emph {et~al.}(2016)\citenamefont {Ni},
  \citenamefont {Jiang},\ and\ \citenamefont {Bambi}}]{Ni:2016uik}%
  \BibitemOpen
  \bibfield  {author} {\bibinfo {author} {\bibfnamefont {Y.}~\bibnamefont
  {Ni}}, \bibinfo {author} {\bibfnamefont {J.}~\bibnamefont {Jiang}}, \ and\
  \bibinfo {author} {\bibfnamefont {C.}~\bibnamefont {Bambi}},\ }\href
  {\doibase 10.1088/1475-7516/2016/09/014} {\bibfield  {journal} {\bibinfo
  {journal} {JCAP}\ }\textbf {\bibinfo {volume} {09}},\ \bibinfo {pages} {014}
  (\bibinfo {year} {2016})},\ \Eprint {http://arxiv.org/abs/1607.04893}
  {arXiv:1607.04893 [gr-qc]} \BibitemShut {NoStop}%
\bibitem [{\citenamefont {Zinhailo}(2018)}]{Zinhailo:2018ska}%
  \BibitemOpen
  \bibfield  {author} {\bibinfo {author} {\bibfnamefont {A.~F.}\ \bibnamefont
  {Zinhailo}},\ }\href {\doibase 10.1140/epjc/s10052-018-6467-8} {\bibfield
  {journal} {\bibinfo  {journal} {Eur. Phys. J. C}\ }\textbf {\bibinfo {volume}
  {78}},\ \bibinfo {pages} {992} (\bibinfo {year} {2018})},\ \Eprint
  {http://arxiv.org/abs/1809.03913} {arXiv:1809.03913 [gr-qc]} \BibitemShut
  {NoStop}%
\bibitem [{\citenamefont {Paul}(2024)}]{Paul:2023eep}%
  \BibitemOpen
  \bibfield  {author} {\bibinfo {author} {\bibfnamefont {P.}~\bibnamefont
  {Paul}},\ }\href {\doibase 10.1140/epjc/s10052-024-12563-6} {\bibfield
  {journal} {\bibinfo  {journal} {Eur. Phys. J. C}\ }\textbf {\bibinfo {volume}
  {84}},\ \bibinfo {pages} {218} (\bibinfo {year} {2024})},\ \Eprint
  {http://arxiv.org/abs/2312.16479} {arXiv:2312.16479 [gr-qc]} \BibitemShut
  {NoStop}%
\bibitem [{\citenamefont {Konoplya}\ \emph
  {et~al.}(2020{\natexlab{b}})\citenamefont {Konoplya}, \citenamefont
  {Pappas},\ and\ \citenamefont {Stuchlík}}]{Konoplya:2020kqb}%
  \BibitemOpen
  \bibfield  {author} {\bibinfo {author} {\bibfnamefont {R.~A.}\ \bibnamefont
  {Konoplya}}, \bibinfo {author} {\bibfnamefont {T.~D.}\ \bibnamefont
  {Pappas}}, \ and\ \bibinfo {author} {\bibfnamefont {Z.}~\bibnamefont
  {Stuchlík}},\ }\href {\doibase 10.1103/PhysRevD.102.084043} {\bibfield
  {journal} {\bibinfo  {journal} {Phys. Rev. D}\ }\textbf {\bibinfo {volume}
  {102}},\ \bibinfo {pages} {084043} (\bibinfo {year} {2020}{\natexlab{b}})},\
  \Eprint {http://arxiv.org/abs/2007.14860} {arXiv:2007.14860 [gr-qc]}
  \BibitemShut {NoStop}%
\bibitem [{\citenamefont {Rostworowski}(2007)}]{Rostworowski:2006bp}%
  \BibitemOpen
  \bibfield  {author} {\bibinfo {author} {\bibfnamefont {A.}~\bibnamefont
  {Rostworowski}},\ }\href@noop {} {\bibfield  {journal} {\bibinfo  {journal}
  {Acta Phys. Polon. B}\ }\textbf {\bibinfo {volume} {38}},\ \bibinfo {pages}
  {81} (\bibinfo {year} {2007})},\ \Eprint {http://arxiv.org/abs/gr-qc/0606110}
  {arXiv:gr-qc/0606110} \BibitemShut {NoStop}%
\bibitem [{\citenamefont {Nollert}(1993)}]{Nollert:1993zz}%
  \BibitemOpen
  \bibfield  {author} {\bibinfo {author} {\bibfnamefont {H.-P.}\ \bibnamefont
  {Nollert}},\ }\href {\doibase 10.1103/PhysRevD.47.5253} {\bibfield  {journal}
  {\bibinfo  {journal} {Phys. Rev. D}\ }\textbf {\bibinfo {volume} {47}},\
  \bibinfo {pages} {5253} (\bibinfo {year} {1993})}\BibitemShut {NoStop}%
\bibitem [{\citenamefont {Zhidenko}(2006)}]{Zhidenko:2006rs}%
  \BibitemOpen
  \bibfield  {author} {\bibinfo {author} {\bibfnamefont {A.}~\bibnamefont
  {Zhidenko}},\ }\href {\doibase 10.1103/PhysRevD.74.064017} {\bibfield
  {journal} {\bibinfo  {journal} {Phys. Rev. D}\ }\textbf {\bibinfo {volume}
  {74}},\ \bibinfo {pages} {064017} (\bibinfo {year} {2006})},\ \Eprint
  {http://arxiv.org/abs/gr-qc/0607133} {arXiv:gr-qc/0607133} \BibitemShut
  {NoStop}%
\bibitem [{\citenamefont {Iyer}\ and\ \citenamefont
  {Will}(1987)}]{Iyer:1986np}%
  \BibitemOpen
  \bibfield  {author} {\bibinfo {author} {\bibfnamefont {S.}~\bibnamefont
  {Iyer}}\ and\ \bibinfo {author} {\bibfnamefont {C.~M.}\ \bibnamefont
  {Will}},\ }\href {\doibase 10.1103/PhysRevD.35.3621} {\bibfield  {journal}
  {\bibinfo  {journal} {Phys. Rev. D}\ }\textbf {\bibinfo {volume} {35}},\
  \bibinfo {pages} {3621} (\bibinfo {year} {1987})}\BibitemShut {NoStop}%
\bibitem [{\citenamefont {Konoplya}(2003)}]{Konoplya:2003ii}%
  \BibitemOpen
  \bibfield  {author} {\bibinfo {author} {\bibfnamefont {R.~A.}\ \bibnamefont
  {Konoplya}},\ }\href {\doibase 10.1103/PhysRevD.68.024018} {\bibfield
  {journal} {\bibinfo  {journal} {Phys. Rev. D}\ }\textbf {\bibinfo {volume}
  {68}},\ \bibinfo {pages} {024018} (\bibinfo {year} {2003})},\ \Eprint
  {http://arxiv.org/abs/gr-qc/0303052} {arXiv:gr-qc/0303052} \BibitemShut
  {NoStop}%
\bibitem [{\citenamefont {Konoplya}(2004)}]{Konoplya:2004ip}%
  \BibitemOpen
  \bibfield  {author} {\bibinfo {author} {\bibfnamefont {R.~A.}\ \bibnamefont
  {Konoplya}},\ }\href@noop {} {\bibfield  {journal} {\bibinfo  {journal} {J.
  Phys. Stud.}\ }\textbf {\bibinfo {volume} {8}},\ \bibinfo {pages} {93}
  (\bibinfo {year} {2004})}\BibitemShut {NoStop}%
\bibitem [{\citenamefont {Matyjasek}\ and\ \citenamefont
  {Opala}(2017)}]{Matyjasek:2017psv}%
  \BibitemOpen
  \bibfield  {author} {\bibinfo {author} {\bibfnamefont {J.}~\bibnamefont
  {Matyjasek}}\ and\ \bibinfo {author} {\bibfnamefont {M.}~\bibnamefont
  {Opala}},\ }\href {\doibase 10.1103/PhysRevD.96.024011} {\bibfield  {journal}
  {\bibinfo  {journal} {Phys. Rev. D}\ }\textbf {\bibinfo {volume} {96}},\
  \bibinfo {pages} {024011} (\bibinfo {year} {2017})},\ \Eprint
  {http://arxiv.org/abs/1704.00361} {arXiv:1704.00361 [gr-qc]} \BibitemShut
  {NoStop}%
\bibitem [{\citenamefont {Konoplya}(2021{\natexlab{a}})}]{Konoplya:2020fwg}%
  \BibitemOpen
  \bibfield  {author} {\bibinfo {author} {\bibfnamefont {R.~A.}\ \bibnamefont
  {Konoplya}},\ }\href {\doibase 10.1103/PhysRevD.103.044033} {\bibfield
  {journal} {\bibinfo  {journal} {Phys. Rev. D}\ }\textbf {\bibinfo {volume}
  {103}},\ \bibinfo {pages} {044033} (\bibinfo {year} {2021}{\natexlab{a}})},\
  \Eprint {http://arxiv.org/abs/2012.13020} {arXiv:2012.13020 [gr-qc]}
  \BibitemShut {NoStop}%
\bibitem [{\citenamefont {Konoplya}(2021{\natexlab{b}})}]{Konoplya:2021ube}%
  \BibitemOpen
  \bibfield  {author} {\bibinfo {author} {\bibfnamefont {R.~A.}\ \bibnamefont
  {Konoplya}},\ }\href {\doibase 10.1016/j.physletb.2021.136734} {\bibfield
  {journal} {\bibinfo  {journal} {Phys. Lett. B}\ }\textbf {\bibinfo {volume}
  {823}},\ \bibinfo {pages} {136734} (\bibinfo {year} {2021}{\natexlab{b}})},\
  \Eprint {http://arxiv.org/abs/2109.01640} {arXiv:2109.01640 [gr-qc]}
  \BibitemShut {NoStop}%
\bibitem [{\citenamefont {Bolokhov}(2024{\natexlab{c}})}]{Bolokhov:2023dxq}%
  \BibitemOpen
  \bibfield  {author} {\bibinfo {author} {\bibfnamefont {S.~V.}\ \bibnamefont
  {Bolokhov}},\ }\href {\doibase 10.1016/j.physletb.2024.138879} {\bibfield
  {journal} {\bibinfo  {journal} {Phys. Lett. B}\ }\textbf {\bibinfo {volume}
  {856}},\ \bibinfo {pages} {138879} (\bibinfo {year} {2024}{\natexlab{c}})},\
  \Eprint {http://arxiv.org/abs/2310.12326} {arXiv:2310.12326 [gr-qc]}
  \BibitemShut {NoStop}%
\bibitem [{\citenamefont {Skvortsova}(2024)}]{Skvortsova:2024atk}%
  \BibitemOpen
  \bibfield  {author} {\bibinfo {author} {\bibfnamefont {M.}~\bibnamefont
  {Skvortsova}},\ }\href {\doibase 10.1002/prop.202400132} {\bibfield
  {journal} {\bibinfo  {journal} {Fortsch. Phys.}\ }\textbf {\bibinfo {volume}
  {72}},\ \bibinfo {pages} {2400132} (\bibinfo {year} {2024})},\ \Eprint
  {http://arxiv.org/abs/2405.06390} {arXiv:2405.06390 [gr-qc]} \BibitemShut
  {NoStop}%
\bibitem [{\citenamefont {Malik}(2025)}]{Malik:2024cgb}%
  \BibitemOpen
  \bibfield  {author} {\bibinfo {author} {\bibfnamefont {Z.}~\bibnamefont
  {Malik}},\ }\href {\doibase 10.1088/1475-7516/2025/04/042} {\bibfield
  {journal} {\bibinfo  {journal} {JCAP}\ }\textbf {\bibinfo {volume} {04}},\
  \bibinfo {pages} {042} (\bibinfo {year} {2025})},\ \Eprint
  {http://arxiv.org/abs/2412.19443} {arXiv:2412.19443 [gr-qc]} \BibitemShut
  {NoStop}%
\bibitem [{\citenamefont {Bolokhov}(2024{\natexlab{d}})}]{Bolokhov:2024ixe}%
  \BibitemOpen
  \bibfield  {author} {\bibinfo {author} {\bibfnamefont {S.~V.}\ \bibnamefont
  {Bolokhov}},\ }\href {\doibase 10.1140/epjc/s10052-024-12990-5} {\bibfield
  {journal} {\bibinfo  {journal} {Eur. Phys. J. C}\ }\textbf {\bibinfo {volume}
  {84}},\ \bibinfo {pages} {634} (\bibinfo {year} {2024}{\natexlab{d}})},\
  \Eprint {http://arxiv.org/abs/2404.09364} {arXiv:2404.09364 [gr-qc]}
  \BibitemShut {NoStop}%
\bibitem [{\citenamefont {Skvortsova}(2025)}]{Skvortsova:2024eqi}%
  \BibitemOpen
  \bibfield  {author} {\bibinfo {author} {\bibfnamefont {M.}~\bibnamefont
  {Skvortsova}},\ }\href {\doibase 10.1209/0295-5075/adaee2} {\bibfield
  {journal} {\bibinfo  {journal} {EPL}\ }\textbf {\bibinfo {volume} {149}},\
  \bibinfo {pages} {59001} (\bibinfo {year} {2025})},\ \Eprint
  {http://arxiv.org/abs/2503.03650} {arXiv:2503.03650 [gr-qc]} \BibitemShut
  {NoStop}%
\bibitem [{\citenamefont
  {Lütfüoğlu}(2025{\natexlab{b}})}]{Lutfuoglu:2025hwh}%
  \BibitemOpen
  \bibfield  {author} {\bibinfo {author} {\bibfnamefont {B.~C.}\ \bibnamefont
  {Lütfüoğlu}},\ }\href {\doibase 10.1088/1475-7516/2025/06/057} {\bibfield
  {journal} {\bibinfo  {journal} {JCAP}\ }\textbf {\bibinfo {volume} {06}},\
  \bibinfo {pages} {057} (\bibinfo {year} {2025}{\natexlab{b}})},\ \Eprint
  {http://arxiv.org/abs/2504.09323} {arXiv:2504.09323 [gr-qc]} \BibitemShut
  {NoStop}%
\bibitem [{\citenamefont {Dubinsky}(2025{\natexlab{b}})}]{Dubinsky:2025fwv}%
  \BibitemOpen
  \bibfield  {author} {\bibinfo {author} {\bibfnamefont {A.}~\bibnamefont
  {Dubinsky}},\ }\href
  {{https://www.sciltp.com/journals/ijgtp/articles/2506000753}} {\bibfield
  {journal} {\bibinfo  {journal} {{International Journal of Gravitation and
  Theoretical Physics}}\ }\textbf {\bibinfo {volume} {1}},\ \bibinfo {pages}
  {2} (\bibinfo {year} {2025}{\natexlab{b}})},\ \Eprint
  {http://arxiv.org/abs/2507.00256} {arXiv:2507.00256 [gr-qc]} \BibitemShut
  {NoStop}%
\bibitem [{\citenamefont {Konoplya}\ \emph {et~al.}(2019)\citenamefont
  {Konoplya}, \citenamefont {Zhidenko},\ and\ \citenamefont
  {Zinhailo}}]{Konoplya:2019hlu}%
  \BibitemOpen
  \bibfield  {author} {\bibinfo {author} {\bibfnamefont {R.~A.}\ \bibnamefont
  {Konoplya}}, \bibinfo {author} {\bibfnamefont {A.}~\bibnamefont {Zhidenko}},
  \ and\ \bibinfo {author} {\bibfnamefont {A.~F.}\ \bibnamefont {Zinhailo}},\
  }\href {\doibase 10.1088/1361-6382/ab2e25} {\bibfield  {journal} {\bibinfo
  {journal} {Class. Quant. Grav.}\ }\textbf {\bibinfo {volume} {36}},\ \bibinfo
  {pages} {155002} (\bibinfo {year} {2019})},\ \Eprint
  {http://arxiv.org/abs/1904.10333} {arXiv:1904.10333 [gr-qc]} \BibitemShut
  {NoStop}%
\bibitem [{\citenamefont {Konoplya}\ and\ \citenamefont
  {Stuchlík}(2017)}]{Konoplya:2017wot}%
  \BibitemOpen
  \bibfield  {author} {\bibinfo {author} {\bibfnamefont {R.~A.}\ \bibnamefont
  {Konoplya}}\ and\ \bibinfo {author} {\bibfnamefont {Z.}~\bibnamefont
  {Stuchlík}},\ }\href {\doibase 10.1016/j.physletb.2017.06.015} {\bibfield
  {journal} {\bibinfo  {journal} {Phys. Lett. B}\ }\textbf {\bibinfo {volume}
  {771}},\ \bibinfo {pages} {597} (\bibinfo {year} {2017})},\ \Eprint
  {http://arxiv.org/abs/1705.05928} {arXiv:1705.05928 [gr-qc]} \BibitemShut
  {NoStop}%
\bibitem [{\citenamefont {Konoplya}\ \emph {et~al.}(2022)\citenamefont
  {Konoplya}, \citenamefont {Zinhailo}, \citenamefont {Kunz}, \citenamefont
  {Stuchlik},\ and\ \citenamefont {Zhidenko}}]{Konoplya:2022hll}%
  \BibitemOpen
  \bibfield  {author} {\bibinfo {author} {\bibfnamefont {R.~A.}\ \bibnamefont
  {Konoplya}}, \bibinfo {author} {\bibfnamefont {A.~F.}\ \bibnamefont
  {Zinhailo}}, \bibinfo {author} {\bibfnamefont {J.}~\bibnamefont {Kunz}},
  \bibinfo {author} {\bibfnamefont {Z.}~\bibnamefont {Stuchlik}}, \ and\
  \bibinfo {author} {\bibfnamefont {A.}~\bibnamefont {Zhidenko}},\ }\href
  {\doibase 10.1088/1475-7516/2022/10/091} {\bibfield  {journal} {\bibinfo
  {journal} {JCAP}\ }\textbf {\bibinfo {volume} {10}},\ \bibinfo {pages} {091}
  (\bibinfo {year} {2022})},\ \Eprint {http://arxiv.org/abs/2206.14714}
  {arXiv:2206.14714 [gr-qc]} \BibitemShut {NoStop}%
\bibitem [{\citenamefont {Konoplya}\ and\ \citenamefont
  {Zhidenko}(2024{\natexlab{b}})}]{Konoplya:2023kem}%
  \BibitemOpen
  \bibfield  {author} {\bibinfo {author} {\bibfnamefont {R.~A.}\ \bibnamefont
  {Konoplya}}\ and\ \bibinfo {author} {\bibfnamefont {A.}~\bibnamefont
  {Zhidenko}},\ }\href {\doibase 10.1103/PhysRevD.109.043014} {\bibfield
  {journal} {\bibinfo  {journal} {Phys. Rev. D}\ }\textbf {\bibinfo {volume}
  {109}},\ \bibinfo {pages} {043014} (\bibinfo {year} {2024}{\natexlab{b}})},\
  \Eprint {http://arxiv.org/abs/2310.19205} {arXiv:2310.19205 [gr-qc]}
  \BibitemShut {NoStop}%
\bibitem [{\citenamefont {Giesler}\ \emph {et~al.}(2019)\citenamefont
  {Giesler}, \citenamefont {Isi}, \citenamefont {Scheel},\ and\ \citenamefont
  {Teukolsky}}]{Giesler:2019uxc}%
  \BibitemOpen
  \bibfield  {author} {\bibinfo {author} {\bibfnamefont {M.}~\bibnamefont
  {Giesler}}, \bibinfo {author} {\bibfnamefont {M.}~\bibnamefont {Isi}},
  \bibinfo {author} {\bibfnamefont {M.~A.}\ \bibnamefont {Scheel}}, \ and\
  \bibinfo {author} {\bibfnamefont {S.}~\bibnamefont {Teukolsky}},\ }\href
  {\doibase 10.1103/PhysRevX.9.041060} {\bibfield  {journal} {\bibinfo
  {journal} {Phys. Rev. X}\ }\textbf {\bibinfo {volume} {9}},\ \bibinfo {pages}
  {041060} (\bibinfo {year} {2019})},\ \Eprint
  {http://arxiv.org/abs/1903.08284} {arXiv:1903.08284 [gr-qc]} \BibitemShut
  {NoStop}%
\bibitem [{\citenamefont {Giesler}\ \emph {et~al.}(2025)\citenamefont {Giesler}
  \emph {et~al.}}]{Giesler:2024hcr}%
  \BibitemOpen
  \bibfield  {author} {\bibinfo {author} {\bibfnamefont {M.}~\bibnamefont
  {Giesler}} \emph {et~al.},\ }\href {\doibase 10.1103/PhysRevD.111.084041}
  {\bibfield  {journal} {\bibinfo  {journal} {Phys. Rev. D}\ }\textbf {\bibinfo
  {volume} {111}},\ \bibinfo {pages} {084041} (\bibinfo {year} {2025})},\
  \Eprint {http://arxiv.org/abs/2411.11269} {arXiv:2411.11269 [gr-qc]}
  \BibitemShut {NoStop}%
\end{thebibliography}%
\end{document}